\documentclass[aps,prd,twocolumn,superscriptaddress,amssymb,eqsecnum,showpacs,showkeyes,secnumarabic,graphics,floatfix,nofootinbib,tightenlines,longbibliography]{revtex4-1}

\usepackage{graphicx}
\usepackage{epsfig}
\usepackage{bm}
\usepackage{cancel}
\usepackage{dcolumn}
\usepackage{amsmath}
\usepackage{hhline}
\usepackage{multirow}
\usepackage{longtable}
\usepackage[utf8]{inputenc}
\usepackage[breaklinks=true,colorlinks=true,
linkcolor=blue,urlcolor=blue,citecolor=cyan,
bookmarks=true,bookmarksopenlevel=2]{hyperref}

\def \beq  {\begin{equation}}
\def \eeq  {\end{equation}}
\def \ber  {\begin{eqnarray}}
\def \eer  {\end{eqnarray}}

\begin{document}
\newcommand{\newc}{\newcommand}

\newc{\be}{\begin{equation}}
\newc{\ee}{\end{equation}}
\newc{\ba}{\begin{eqnarray}}
\newc{\ea}{\end{eqnarray}}
\newc{\bea}{\begin{eqnarray*}}
\newc{\eea}{\end{eqnarray*}}
\newc{\D}{\partial}
\newc{\ie}{{\it i.e.} }
\newc{\eg}{{\it e.g.} }
\newc{\etc}{{\it etc.} }
\newc{\etal}{{\it et al.}}
\newc{\lcdm}{$\Lambda$CDM }
\newcommand{\nn}{\nonumber}
\newc{\ra}{\Rightarrow}
\newc{\lsim}{\buildrel{<}\over{\sim}}
\newc{\gsim}{\buildrel{>}\over{\sim}}

\title{Sub-millimeter Spatial Oscillations of Newton's Constant: Theoretical Models and Laboratory Tests}

\author{ L. Perivolaropoulos} \email{leandros@uoi.gr} 
\affiliation{Department of Physics, University of Patras, 26500 Patras, Greece \\(on leave from the Department of Physics, University of Ioannina, 45110 Ioannina, Greece) }

\date {\today}

\begin{abstract}
We investigate the viability of sub-millimeter wavelength oscillating deviations from the Newtonian potential at both the theoretical and the experimental/observational level. At the theoretical level such deviations are generic predictions in a wide range of extensions of General Relativity (GR) including $f(R)$ theories, massive Brans-Dicke (BD)- scalar tensor theories, compactified extra dimension models and nonlocal extensions of GR. 
However, the range of parameters associated with such oscillating deviations is usually connected with instabilities present at the perturbative level. An important exception emerges in nonlocal gravity theories where oscillating deviations from Newtonian potential occur naturally on sub-millimeter scales without any instabilities. As an example of a model with unstable Newtonian oscillations we review an $f(R)$ expansion around General Relativity of the form  $f(R)=R+\frac{1}{6 m^2} R^2$ with $m^2<0$ pointing out possible stabilization mechanisms. As an example of a model with stable Newtonian oscillations we discuss nonlocal gravity theories.
If such oscillations are realized in Nature on sub-millimeter scales, a signature is expected in torsion balance experiments testing the validity of Newton's law. We search for such a signature in the torsion balance data of the Washington experiment \cite{Kapner:2006si-washington3} (combined torque residuals of experiments I, II, III) testing Newton's law at sub-millimeter scales. We show that an oscillating residual ansatz with spatial wavelength $\lambda \simeq 0.1mm$ provides a better fit to the data compared to the residual Newtonian constant ansatz by $\Delta \chi^2 = -15$. Similar improved fits however, also occur in about $10\%$ of Monte Carlo realization of Newtonian data on similar or larger scales. Thus, the significance level of this improved fit is at a level of not more than $2\sigma$. The energy scale corresponding to this best fit wavelength is identical to the dark energy length scale $\lambda_{de} \equiv\sqrt[4]{\hbar c/\rho_{ de}}\approx  0.1mm$. 

\end{abstract}
\maketitle

\section{Introduction}
\label{sec:Introduction}

The discovery of the accelerating expansion of the universe\citep{Riess:1998cb,Peebles:2002gy,Caldwell:2009ix-review-accel-univ} has opened a new prospect for a possible need for modification of General Relativity beyond the level of a cosmological constant. Such a geometric origin of dark energy is simple and well motivated physically\cite{Copeland:2006wr-dark-energy-review,Clifton:2011jh-gen-mod-grav-cosmo-rev1,Nojiri:2008nt-review-dark-energy-from-fR,Amendola:2016saw-review-dark-energy-fR-comments,Capozziello:2011et-gen-mod-grav-rev}.

The typical physics scale of such geometric dark energy required so that it starts dominating the universe at recent cosmological times is $\lambda_{de} \equiv\sqrt[4]{\hbar c/\rho_{ de}}\approx  0.085 mm$ (assuming $\Omega_{0m}=0.3$ and $H_0=70 km sec^{-1} Mpc^{-1}$). Therefore, if the origin of the accelerating expansion is geometrical, it is natural to expect the presence of signatures of modified gravity on scales of about $0.1 mm$. 

A wide range of experiments has focused on this range of scales \citep{Murata:2014nra-good-review-ofexperiments,Kapner:2006si-washington3,Hoyle:2004cw-washington2,Hoyle:2000cv-washington1} and constraints have been imposed on particular parametrizations of extensions of Newton's gravitational potential. Such parametrizations are motivated by viable extensions of General Relativity and include Yukawa interactions leading to an effective gravitational potential
\be 
V_{eff}= -G \frac{M}{r}(1+\alpha e^{- m r})
\label{yukawaanz}
\ee
and a power-law ansatz of the form \cite{Adelberger:2006dh-yukawa-power-law-constraints}
\be 
V_{eff}= -G \frac{M}{r}(1+\beta^k(\frac{1}{m r})^{k-1})
\label{poweranz}
\ee
arising for example in the context of some brane world models\cite{Donini:2016kgu,Benichou:2011dx,Bronnikov:2006jy,Nojiri:2002wn-newtonptl-brane}. 

The Yukawa interaction parametrization (\ref{yukawaanz}) is motivated by the weak gravitational field limit solution of a point mass in a wide range of extensions of GR including $f(R)$ theories\citep{Berry:2011pb-weak-field-fr-incl-osc,Capozziello:2009vr-fR-newtonian-limit-nooscil,Schellstede:2016ldu-newtonian-fR-oscillations} massive Brans-Dicke (BD)\cite{Perivolaropoulos:2009ak-massive-bd-ppn,Hohmann:2013rba-gamma-par-scal-tens1,Jarv:2014hma-gamma-par-scal-tens2}  and scalar tensor theories, compactified extra dimension models \cite{ArkaniHamed:1998rs,ArkaniHamed:1998nn,Antoniadis:1998ig,Perivolaropoulos:2002pn-radion-yukawa,Floratos:1999bv-radion-yukawa,Kehagias:1999my} etc. In each of these models the mass scale $m$ has a different physical origin. For example in massive BD theories $m$ is the mass scale of the BD scalar and $\alpha=\frac{1}{3+2\omega}$ while in $f(R)$ theories it is the lowest order term series expansion of $f(R)$ around GR of the form
\be
f(R)=R+\frac{1}{6m^2}R^2 + ...
\label{frexp}
\ee
In $f(R)$ theories we have $\alpha=\frac{1}{3}$\cite{Capozziello:2007ms-fR-newtonian-limit-with-osc-good-disc}. In radion compactified extra dimension models $m$ is the inverse radius of the extra dimension ($m=b^{-1}$) while $\alpha=\frac{D}{D+2}$ \cite{Perivolaropoulos:2002pn-radion-yukawa} where $D$ is the number of toroidally compactified dimensions even though different values may be obtained for different compactifications \cite{Kehagias:1999my}. 

In all these theories, the weak field limit solution remains mathematically valid and consistent for $m^2 <0$\cite{Berry:2011pb-weak-field-fr-incl-osc,Capozziello:2007ms-fR-newtonian-limit-with-osc-good-disc}. For this mass range the correction of the effective gravitational potential becomes oscillating takes the form
\be 
V_{eff}= -G \frac{M}{r}(1+\alpha \cos(m r+\theta))
\label{oscillanz}
\ee
where $\theta$ is an arbitrary parameter.
Such a potential clearly has no Newtonian limit and could be ruled out immediately on this basis\cite{Olmo:2005hc-fR-newtonian-osc-discussed-ruledout-no-instab-disc,Schellstede:2016ldu-newtonian-fR-oscillations}. However, since the extra force component averages out to zero, for sub-millimeter wavelength oscillations and $\alpha=O(1)$ such an oscillating correction could remain undetectable by current experiments and astrophysical observations due to finite accuracy in length and force measurements. 

This class of theories however, has an additional problem: the vacuum in most of such theories suffers from serious instabilities\cite{Dolgov:2003px-1overR-term-instability,Faraoni:2006sy-fR-stability-good} (see also \cite{Nojiri:2003ft}) in this range of values of $m^2$. Due to these problems, the oscillating correction has not been considered in any detail and there are currently no experimental constraints on the corresponding parametrization parameters.

A peculiar feature of these perturbative results\cite{Faraoni:2006sy-fR-stability-good,Dolgov:2003px-1overR-term-instability,Capozziello:2009nq-fR-with-neg-alpha-unstable-dolgov-instability,Faraoni:2007yn-physical-meaning-ofinstability,Navarro:2006mw-fR-newtonian-instability-discussion,Faraoni:2008ke-instability-with-geg-m2,Nojiri:2007jr-instab,Cognola:2007zu-viable-models} is that they indicate the presence of an infinite discontinuity in the stability properties of these theories as $m^{-2}\rightarrow 0^-$. In this limit $f(R)$ theories are extremely unstable (with lifetime of the vacuum that approaches zero as $m^{-2}\rightarrow 0$). Exactly at $m^{-2}=0$ however, the theory becomes stable and identical to GR. The existence of this unphysical infinite discontinuity may indicate that non-perturbative effects and non-trivial backgrounds may play a significant role in the stability analysis. Such effects are briefly discussed in the present analysis.

A more promising theoretical model where stable spatial oscillations naturally occur for the gravitational potential are non-local theories of gravity\cite{Edholm:2016hbt-nonlocal-potential-stable-spatial-oscillations,Kehagias:2014sda-nonlocal-oscillations,Frolov:2015usa-newton-potential-nonlocal} motivated from string theory and described by actions that are made generally covariant and ghost free at the perturbative level by including infinite derivatives  \cite{Tomboulis:1997gg,Siegel:2003vt,Biswas:2013cha}. For example such an action, viable also at the cosmological level is of the form \cite{Dirian:2016puz-acceleration-from-nonlocal-gravity}
\be
S=\frac{1}{16\pi G}\int d^4 x\sqrt{-g}\left[R+\frac{1}{6 m^2}R\left(1-\frac{\Lambda^4}{\Box^2}\right)R\right]
\label{nonlocalaction}
\ee
where $m$ is the scale of non-locality and $\Box$ is the d' Alembert operator. 

In this class of theories the generalized Newtonian potential remains finite at $r=0$ and may develop spatial oscillations\cite{Edholm:2016hbt-nonlocal-potential-stable-spatial-oscillations,Kehagias:2014sda-nonlocal-oscillations,Maggiore:2014sia-nonlocal-gravity-oscil} around the scale of non-locality (see section II-c) which decay on larger scales. Despite of increased complexity this class of theories has four important advantages
\begin{itemize}
\item
They can be free from singularities while having a proper Newtonian limit \cite{Frolov:2015usa-newton-potential-nonlocal}.
\item 
They are free from instabilities in the absence of tensorial nonlocal terms \cite{Nersisyan:2016jta-nonlocal-instabilities}.
\item
They can emerge from effects at the quantum level. In particular, light particle loops at the quantum
level can generate non-local terms in the quantum
effective action which can make it renormalizable \cite{Talaganis:2014ida}
\item
They are consistent with the cosmological observations without need for cosmological constant \cite{Dirian:2016puz-acceleration-from-nonlocal-gravity,Park:2012cp-non-local-gravity-geff-cosmology,Calcagni:2010ab-nonlocal-gravity-cosmology,Barvinsky:2011hd-nonlocal-gravity-cosmology}.
\end{itemize}
Thus, in view of the generic nature of oscillating parametrizations and the fact that there may be stabilization mechanisms like backreaction from higher order nonlinear or nonlocal terms, it is of interest to consider in some detail the theoretical models and the observational consequences of such an oscillating correction of Newton's law potential. This is the goal of the present analysis.

\begin{figure}[!t]
\centering
\vspace{0cm}\rotatebox{0}{\vspace{0cm}\hspace{0cm}\resizebox{0.49\textwidth}{!}{\includegraphics{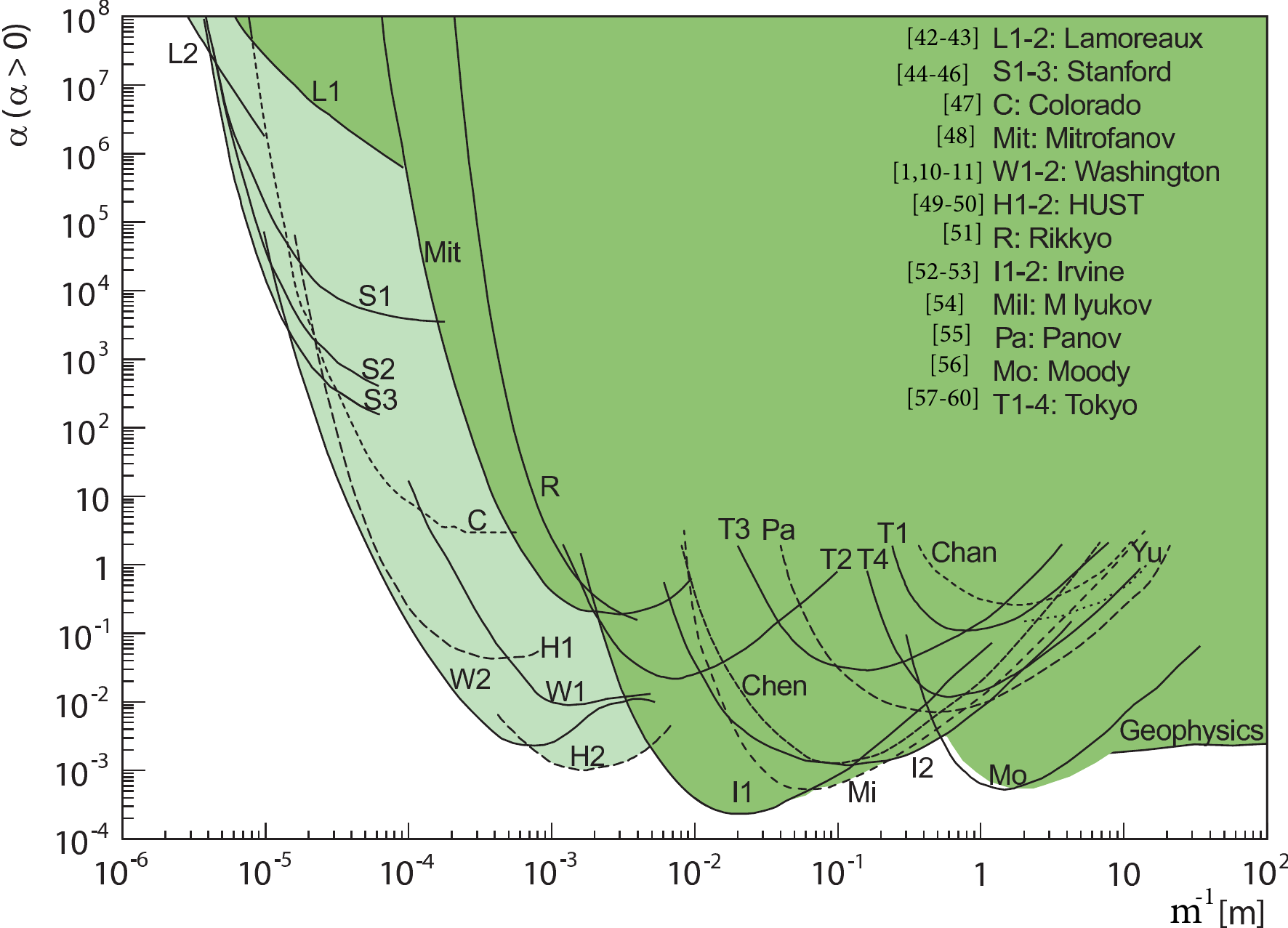}}}
\caption{A review of current constraints\cite{L1-2Lamoreaux,L1-2Sushkov,S1-3Chiaverini1,S1-3Smullin1,S1-3Geraci1,
ColoradoLong1,Mitrofanov1,Kapner:2006si-washington3,Hoyle:2004cw-washington2,Hoyle:2000cv-washington1,H1-2Tu1,H1-2Yang1,RikkyoMurata1,I1-2Hoskins1,I1-2Spero,Milyukov1,Panov1,Moody1,
T1-4Hirakawa1,T1-4Ogawa1,T1-4Kuroda1,T1-4Mio1} 
based on the Yukawa parametrization (\ref{yukawaanz}) for deviation from Newton's law (from Ref. \citep{Murata:2014nra-good-review-ofexperiments}).}
\label{experiments}
\end{figure}

A wide range of experiments (see Fig. \ref{experiments} and Ref. \cite{Murata:2014nra-good-review-ofexperiments} for a good review) have been performed during the recent years imposing constraints on deviations from Newton's law using the Yukawa and the power law parametrizations. These include torsion balance experiments \cite{ColoradoLong1,Panov1,Mitrofanov1,I1-2Hoskins1,I1-2Spero,
Chen1,Chan1,Moody1,T1-4Hirakawa1,T1-4Ogawa1,T1-4Kuroda1,T1-4Mio1,
RikkyoMurata1,S1-3Smullin1,S1-3Geraci1,S1-3Chiaverini1}   measuring gravitational torques from source masses on  test masses
attached to torsion balance bars, Casimir force experiments \cite{L1-2Lamoreaux,L1-2Sushkov} looking for anomalies in the electric forces between two metal surfaces and atomic or nuclear experiments \cite{Murata:2014nra} looking for anomalies in the dynamical evolution of particle systems in the context of known standard model interactions.
The constraints of such experiments on Yukawa deviations from a Newtonian potential are reviewed in Fig. \ref{experiments}. 

The most constraining experiments testing the Yukawa parametrization (\ref{yukawaanz})  for deviations from Newton's law for $\alpha=O(1)$ have been performed using a torsion balance instrument by the Washington group \citep{Kapner:2006si-washington3} in 2006. For $\alpha=1$ the $2\sigma$ constraint on $m$ was obtained as $m\gsim 18 mm^{-1}$. An interesting open question is the following: 'What are corresponding constraints in the context of an oscillating parametrization for deviations from Newton's law?' The answer to this question is one of the main goals of the present analysis.

The structure of this paper is the following: In the next section we discuss the emergence of an oscillating deviation from Newton's constant in a particular class of theoretical modified gravity models: $f(R)$ theories\citep{DeFelice:2010aj-review-fR-theories} with $f(R)=R+\frac{1}{6 m^2} R^2$. We derive the parameter range of $m^2$ for which an oscillating deviation from the Newtonian potential emerges using both the $f(R)$ formalism\citep{Chiba:2006jp-gmonehalf} and the equivalent massive BD formalism\citep{Chiba:2003ir-gamma05-weak-field}. We also discuss the stability of this solution and confirm that it is unstable in accordance with the Dolgov-Kawasaki-Faraoni instability \citep{Dolgov:2003px-1overR-term-instability,Faraoni:2008ke-instability-with-geg-m2}. The gravitational potential that emerges in non-local gravity theories is also discussed and shown to have oscillating deviations from the Newtonian potential. Even though these deviations are in general complicated functional expressions\cite{Edholm:2016hbt-nonlocal-potential-stable-spatial-oscillations}, we show that in appropriate limits they are very well fit by simple trigonometric functions. In section III we discuss the consistency of these submillimeter oscillations of the generalized Newtonian potential with macroscopic large scale observations given the finite accuracy in the measurement of forces and lengths. We also use an oscillating parametrization to fit the torsion balance data of the Washington group experiment. Finally in section IV we conclude, summarise and discuss future extensions of the present analysis. \footnote{In what follows we use the metric signature $(-+++)$. This is important in view of the fact that the sign of the Ricci scalar changes if a different metric signature is used and thus the sign of $m^2$ in eq. (\ref{frexp})  leading to stability or instability would also change.}

\section{Spatial Oscillations of Newton's Constant in modified gravity theories}
\label{sec:Section 2}
\subsection{$f(R)$ theories I: The equivalent BD formalism and the Newtonian limit}

The Einstein-Hilbert action is generalized in the context of $f(R)$ theories as
\be 
S_R=\frac{1}{16 \pi G}\int d^4 x \sqrt{-g} f(R)+S_{matter}
\label{fraction}
\ee
where $R$ is the Ricci scalar. This action may be shown to be dynamically equivalent to the scalar-tensor action of a massive BD scalar field with $\omega=0$\cite{Chiba:2003ir-gamma05-weak-field,Capozziello:2010wt-ohanlon-gravity-and-fR-incorrect-in-Newtonian,Faraoni:2006hx-ill-defined-equivalency-st-fR}
\ba 
S_{BD}&=&\frac{1}{16 \pi G}\int d^4 x \sqrt{-g} \left[ f(\phi)+f_\phi(\phi)(R-\phi)\right]\nn \\ &+&S_{matter}
\label{bdequiv}
\ea
Variation of the action (\ref{bdequiv}) with respect to the field $\phi$ leads to the condition $\phi=R$ (assuming $f_{\phi\phi}\neq 0)$ which reduces the action (\ref{bdequiv}) to the $f(R)$ action (\ref{fraction})\cite{Chiba:2003ir-gamma05-weak-field}.
Assuming an $f(R)$ theory of the form
\ba
f(R)=R&+\frac{1}{6m^2}R^2  
\label{franz} \ea
the scalar field action (\ref{bdequiv}) is easily shown to take the form
\ba
S_{BD}&=&\frac{1}{16 \pi G}\int d^4 x \sqrt{-g} \left[(1+ \frac{1}{3m^2}\phi)R -\frac{1}{6m^2}\phi^2\right]\nn \\&+&S_{matter}
\label{bdequiv2}
\ea
We now define the field $\Phi \equiv 1 + \frac{1}{3m^2}\phi$ and the action (\ref{bdequiv2}) takes the form
\ba
S_{BD}&=&\frac{1}{16 \pi G}\int d^4 x \sqrt{-g} \left[\Phi R -\frac{3}{2}m^2 (\Phi-1)^2\right]\nn \\ &+& S_{matter}
\label{bdequiv3}
\ea
The action (\ref{bdequiv3}) is identical with the action of a massive BD scalar field with $\omega=0$. Due to the non-zero mass of the BD scalar PPN parameter $\gamma$ is not of the form $\gamma=\frac{1+\omega}{2+\omega}=\frac{1}{2}$ as in the massless case.
The Newtonian limit of this theory has been investigated in detail in Ref. \cite{Perivolaropoulos:2009ak-massive-bd-ppn} but we review that analysis here for completeness setting $\omega=0$.

The dynamical field equations obtained by varying the action (\ref{bdequiv3}) are of the form
\begin{widetext}
\be
\Phi \left(R_{\mu\nu}-{1\over2}g_{\mu\nu}R\right)
= 8\pi G T_{\mu\nu}
+\nabla_\mu\partial_\nu \Phi - g_{\mu\nu}\Box \Phi
- g_{\mu\nu} \frac{3}{4}  m^2 (\Phi-1)^2 \label{metdyneq} \ee
\be
\Box\Phi =
\frac{8\pi G}{3} \,T + m^2 \left((\Phi-1)^2 + (\Phi-1)\Phi\right)
\label{phidyneq}
\ee
\end{widetext}
We now consider the weak gravitational field of a point mass with
\be
T_{\mu\nu}=diag(M\delta(\vec r),0,0,0)
\label{tmnpointmass}
\ee
and thus we expand around a constant-uniform background field $\Phi_0=1$ and a Minkowski metric $\eta_{\mu\nu}=diag(-1,1,1,1)$
\ba \Phi&=&1 +\varphi \label{expf}  \\ g_{\mu\nu}&=&\eta_{\mu\nu}+h_{\mu\nu} \label{expg} \ea
The resulting equations for $\varphi$ and $h_{\mu\nu}$ obtained from (\ref{metdyneq}), (\ref{phidyneq}), (\ref{expf})and (\ref{expg}) in the gauge $h^\mu_\nu,_\mu-\frac{1}{2}h^\mu_\mu,_\nu=\varphi,_\nu$ are of the form
\be
\left(\Box -  m^2 \right)\varphi = -\frac{8\pi G}{3} M \delta(\vec r) \label{linphieq} \ee
\be -\frac{1}{2} \left[\Box (h_{\mu\nu}-\eta_{\mu\nu}\frac{h}{2})\right]= 8\pi G T_{\mu\nu} +\partial_\mu\partial_\nu \varphi-\eta_{\mu\nu}\Box\varphi \label{linmeteq} \ee
where $h=h^\mu_\mu$. For static configurations equations (\ref{linphieq}), (\ref{linmeteq}) become \ba \nabla^2\varphi -  m^2 \varphi &=& -\frac{8\pi G}{3}M \delta(\vec r) \label{linphieq1} \\
\nabla^2 h_{00} - \nabla^2 \varphi &=& -8\pi G M \delta(\vec r) \label{linmeteq1a} \\
\nabla^2 h_{ij} - \delta_{ij} \nabla^2 \varphi &=& -8\pi GM \delta(\vec r)  \delta_{ij} \label{linmeteq1b} \ea
These equations have the following solution \ba \varphi &=&\frac{2 G M}{3r}e^{-m r} \label{phisol} \\
h_{00}&=&\frac{2GM}{r}\left(1+\frac{1}{3}e^{-m r}\right)\label{h00sol} \\
h_{ij}&=&\frac{2GM}{r}\delta_{ij}\left(1-\frac{1}{3}e^{-m r}\right)\label{hijsol} \ea 

The metric may now be expanded in terms of the $\gamma$ post-Newtonian parameter as \ba g_{00}&=&-(1+2 V_{eff}) \label{gamdef1} \\ g_{ij}&=&(1-2\gamma V_{eff})\delta_{ij} \label{gamdef2}\ea where $V_{eff}$ is the Newtonian potential. Thus we obtain \be \gamma(m,r)=\frac{h_{ij}\vert_{i=j}}{h_{00}}=\frac{3 -e^{-m r}}{3+e^{-m r}} \label{gamfull} \ee
In the special case of $m=0$ we obtain the result of Ref. \cite{Chiba:2006jp-gmonehalf,Olmo:2006eh-fR-ruled-out-sameas-chiba-1,Erickcek:2006vf-fR-ruled-out-sameas-chiba,Bertolami:2013qaa-sim-to-chiba-drops-exp}  $\gamma=\frac{1}{2}$ (extreme deviation from GR) while for $m\rightarrow \infty$ we recover the GR limit $\gamma\rightarrow 1$. It is therefore clear that these theories are viable and have a well defined Newtonian limit in contrast to the conclusion of some previous studies (eg \cite{Chiba:2006jp-gmonehalf}).

The generalized Newtonian potential takes the form 
\be V_{eff}=-\frac{h_{00}}{2}=-\frac{G M}{r}\left(1+\frac{1}{3}e^{-m r}\right) \label{ntpot1} \ee 
while the corresponding force is
\be F_{eff}=-\hat{r}\frac{G M}{r^2}\left(1+\frac{e^{-m r}}{3}+\frac{m r}{3} e^{-m r}\right)
\label{ntforceexp}
\ee 
The stability of the solution (\ref{phisol})-(\ref{hijsol}) may be studied by considering perturbations of the form $\varphi=\varphi_0(r) + \delta\varphi(r,t)$ where $\varphi_0$ is the unperturbed static solution (\ref{phisol}). The perturbation satisfies the equation
\be
-\ddot \delta\varphi + \nabla^2 \delta\varphi - m^2 \delta\varphi =0
\label{perteq1}
\ee
For $m^2 >0$ this is the Klein-Gordon equation which has only wavelike solutions and therefore $\varphi_0$ is stable. The presence of higher order terms in the Lagrangian (\ref{bdequiv3}) however could change the stability properties of this solution since it is well known that the non-linear Klein-Gordon equation can have instabilities around non-trivial solutions (eg \cite{Dolan:2007mj}). The vacuum solution however ($\varphi=0$) is always stable for $m^2>0$.

For $m^2<0$ the weak field solution of eq. (\ref{linphieq1}) becomes oscillatory and takes the form
\be
\varphi =\frac{2 G M}{3r}\cos(\vert m \vert r+\theta)
\label{phisolnegm2} 
\ee
where $\theta$ is an arbitrary phase. 
The Newtonian potential in this case takes the form 
\be V_{eff}=-\frac{h_{00}}{2}=-\frac{G M}{r}\left(1+\frac{1}{3}\cos(\vert m \vert r+\theta)\right) \label{ntpot2} \ee 

This solution is perturbatively unstable in the absence of higher order terms in the action or in the absence of a nontrivial background energy momentum tensor. In addition, the trivial vacuum solution ($\varphi_0=0$) is clearly always unstable for this range of $m^2$.

\subsection{$f(R)$ theories II: The Newtonian limit in the $f(R)$ formalism}

We now rederive the form of the weak field metric and the effective Newtonian potential using directly the $f(R)$ formalism\citep{Capozziello:2007ms-fR-newtonian-limit-with-osc-good-disc,Capozziello:2009vr-fR-newtonian-limit-nooscil,Castel-Branco:2014kja-newt-limit-exp-only,Chiba:2003ir-gamma05-weak-field,Chiba:2006jp-gmonehalf,Berry:2011pb-weak-field-fr-incl-osc,Olmo:2006eh-fR-ruled-out-sameas-chiba-1} since there has been some controversy in the literature concerning the equivalence of the two formalisms\cite{Faraoni:2006hx-ill-defined-equivalency-st-fR}. 

Variation of the $f(R)$ action (\ref{fraction}) with respect to the metric leads to the generalized Einstein equations
\be  f'(R)R_{\mu\nu} - \frac{1}{2}g_{\mu\nu}f(R) = 8\pi G T_{\mu\nu} + \nabla_{\mu}\nabla_{\nu}f'(R) - g_{\mu\nu}\Box f'(R)
\label{frdyneqs}
\ee
Using the ansatz (\ref{franz})  and the weak field metric expansion (\ref{expg}) while keeping only linear terms in $h_{\mu\nu}$, the dynamical equations takes the form
\be 
 R_{\mu\nu} - \frac{1}{2}R\eta_{\mu\nu} = 8\pi GT_{\mu\nu} + (\partial_\mu \partial_\nu - \eta_{\mu\nu}\nabla^2)\frac{1}{3m^2}R 
\label{dyneqslin}
\ee
Taking the trace of (\ref{dyneqslin})  and using eq. (\ref{tmnpointmass}) we find
\be  \frac{1}{m^2}\nabla^2 R - R = -8\pi G M \delta(\vec r)
\label{tracedyneq1}
\ee
For $m^2>0$ the physically interesting solution of this equation is
\be  \bar R = \frac{2GM}{r} m^2\;e^{-m r}
\label{rsol1}
\ee
while for $m^2<0$ we obtain oscillating form 
\be  \bar R = \frac{2GM}{r} m^2\;\cos(\vert m \vert  r + \theta)
\label{rsol1osc}
\ee
Using now equation (\ref{rsol1}) in eq. (\ref{dyneqslin}) we find
the equation for the $00$ component as

\be  R_{00} = \frac{16}{3}\pi G \rho - \frac{1}{6}R
\label{r00eq}
\ee
At the linear level in $h_{\mu\nu}$ the Ricci tensor is of the form
\be R_{\mu\nu} = \frac{1}{2}(\partial_\sigma\partial_\nu h_\mu^\sigma + \partial_\sigma\partial_\mu h_\nu^\sigma - \partial_\mu\partial_\nu h - \nabla^2 h_{\mu\nu})
\label{rmnlinear}
\ee
which for the $00$ component becomes
\be  
R_{00} = \frac{1}{2}(-\nabla^2h_{00})
\label{rttlinear}
\ee
Using eq. (\ref{rttlinear}) in (\ref{r00eq}) we find
\be 
 \nabla^2h_{00} = -\frac{32}{3}\pi G \rho + \frac{1}{3}R
\label{nablah00}
\ee
Setting $\rho=M \delta({\vec r})$ and using (\ref{rsol1}) in (\ref{nablah00}) we obtain the solution for $h_{00}$ as
\be  h_{00} = \frac{2G M}{r}(1 + \frac{1}{3}e^{-m r})
\label{h00sol1}
\ee
which is identical with the corresponding result (\ref{h00sol}) found using the massive BD formalism. In order to find the $h_{ij}\vert_{i=j}$ components we express $R$ in terms of the linearized metric components as
\be 
 R = -2\nabla^2h_{ij}\vert_{i=j} + \nabla^2h_{00}
\label{rvsh}
\ee
 Using (\ref{nablah00}) in (\ref{rvsh}) we find
\be  \nabla^2h_{ij}\vert_{i=j} = -\frac{16}{3}\pi G\rho - \frac{1}{3}R
\label{nablahii}
\ee
which for  $\rho=M\delta(\vec r)$ and $R$ from eq. (\ref{rsol1}) leads to 

\be  h_{ij} = \frac{2GM}{r}\delta_{ij}(1 - \frac{1}{3}e^{-m r})
\label{hijsol1}
\ee
This is identical to the corresponding result (\ref{hijsol}) obtained using the massive BD formalism. The generalized Newtonian potential and the PPN parameter $\gamma$ are now obtained in exactly the same way as in the massive BD formalism.

For $m^2<0$ the solution reduces to the oscillating form
\ba  h_{00} &=& \frac{2G M}{r}\left(1 + \frac{1}{3}\cos(\vert m \vert r + \theta)\right)\\
h_{ij} &=& \frac{2GM}{r}\delta_{ij}\left(1 - \frac{1}{3}\cos(\vert m \vert r + \theta)\right)
\label{h00sol1osc}
\ea

We thus conclude that the two formalisms are consistent and equivalent and they predict an oscillating correction to the Newtonian potential in accordance with equation (\ref{ntpot2}) for $m^2<0$.

\subsubsection{Stability at the non-linear level}

We now discuss the stability of this oscillating weak field solution in the $f(R)$ formulation in the presence of nonlinear terms. Using the ansatz (\ref{franz}) in the trace of the dynamical equations (\ref{frdyneqs}) we find
\cite{Faraoni:2006sy-fR-stability-good}
\be
-{\ddot R} + \nabla^2 R +\frac{1}{6}R^2 -m^2 R=-8\pi G M m^2 \delta(\vec r)
\label{rdyneqnonlin}
\ee
For large but finite $m^2$ the linear term dominates and this equation reduces to eq. (\ref{tracedyneq1}) with solution (\ref{rsol1})-(\ref{rsol1osc}). 

In order to test the stability of this solution we perturb it setting $R=\bar R + \delta R$ where $\delta R$ is a time dependent perturbation. In view of the fact that the  unperturbed solution $\bar R$ is large ($O(m^2)$) we need to take into account the backreaction from the nonlinear term $\frac{1}{6}R^2$ to find the time evolution of the perturbation $\delta R$. This contribution has not been taken into account in previous stability analyses\cite{Faraoni:2006sy-fR-stability-good}. Thus taking this term into account and keeping only linear terms in $\delta R$ we obtain 
\be 
-\ddot{\delta R}+\nabla^2 \delta R - m^2(1-\frac{1}{3m^2} \bar R)\delta R =0
\label{stabeq1}
\ee
where $\bar R$ is given by eq. (\ref{rsol1}).

Assuming $m^2<0$ and setting $\delta R= \delta R_0(r) e^{\omega t}$ we obtain a Schrodinger-like equation for the perturbation $\delta R_0$ of the form
\be
\left(-\nabla^2 + V(r)\right)\delta R_0 = -\omega^2  \delta R_0
\label{schrod1}
\ee
where the potential is
\be 
V(r)\equiv m^2 \left[1-\frac{r_c}{3r}\cos(\vert m \vert r +\theta)\right]
\label{potential}
\ee
with $r_c\equiv 2 G M$. After a rescaling $\vert m \vert r\rightarrow \bar r$ we are left with a single dimensionless parameter $\bar r_c\equiv \vert m \vert r_c$. The presence of the large nontrivial background solution combined with the presence of the nonlinear term has modified the effective mass by an oscillating $r$-dependent term which is not smaller than the homogeneous mass term. This term could in principle modify the stability properties of the background oscillating solution $\bar R$. Notice however that such a term would not change the stability properties of the vacuum ($\bar R=0$). We have shown by numerical solution of eq. (\ref{schrod1}) that there are unstable mode solutions ($\omega^2>0$) for all values of the dimensionless parameter $\bar r_c$ which rise exponentially with time for finite $\bar{r}$ and go to $0$ at $\bar r\rightarrow \infty$.

\begin{figure}[!t]
\centering
\vspace{0cm}\rotatebox{0}{\vspace{0cm}\hspace{0cm}\resizebox{0.49\textwidth}{!}{\includegraphics{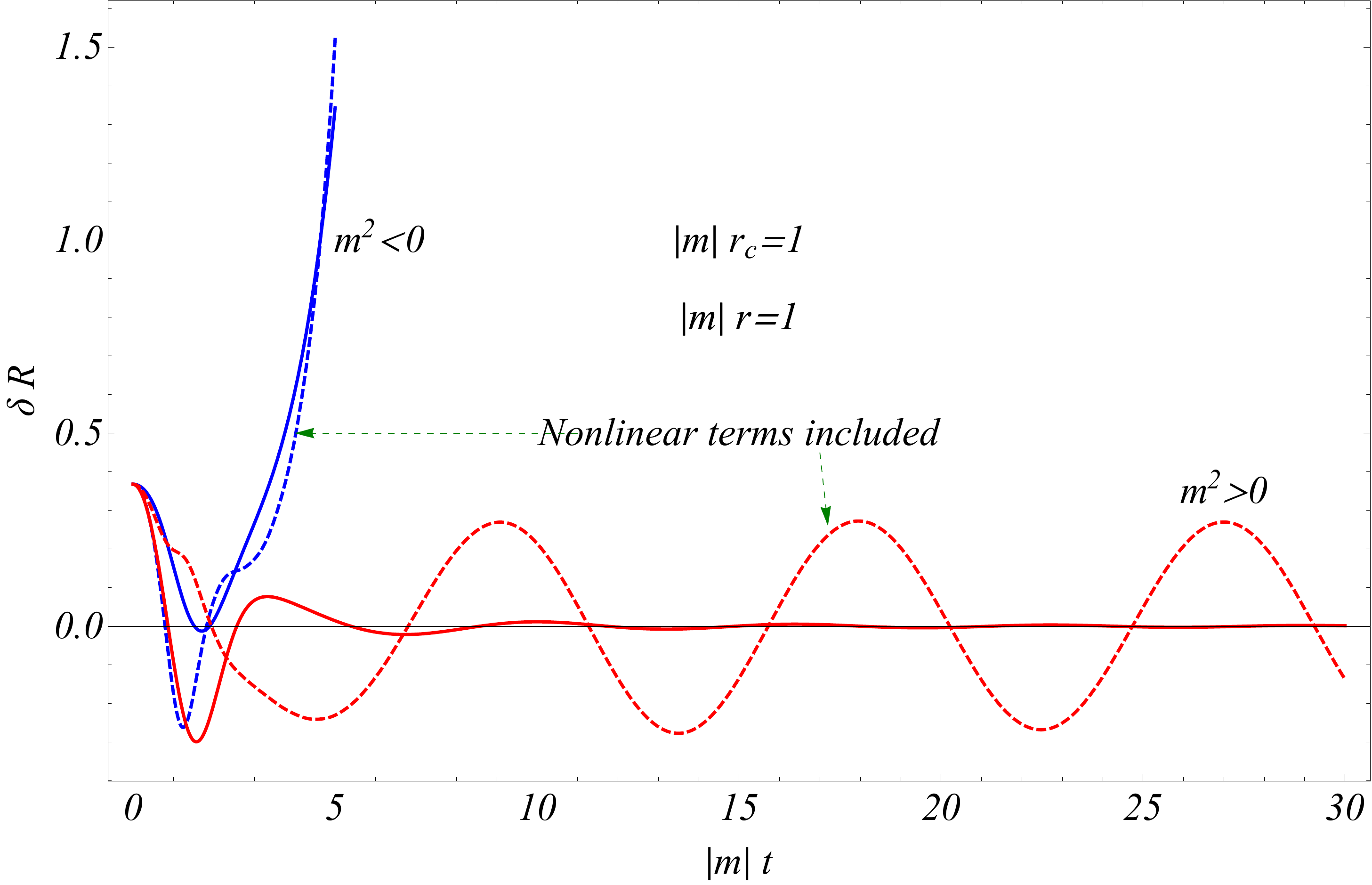}}}
\caption{ The evolution of the Ricci scalar perturbations $\delta R(t,\vert m \vert r=1)$ is shown in  for $m^2>0$ (red lines-stability) and $m^2<0$ (blue lines-instability). The dashed lines correspond to the presence of the nonlinear term with $\bar{r}_c=10$, while for the continous lines the nonlinear term is absent ($\bar{r}_c=0$).}
\label{stability}
\end{figure}

In order to demonstrate the effects of the nonlinear term on the stability of the perturbations we have solved eq. (\ref{stabeq1}) with initial condition $\delta R(t=0,r)= e^{-r^2}$, $\dot{\delta R}(t=0,r)= 0$ for both $m^2>0$ (stability) and $m^2<0$ instability. For $m^2>0$ and proper rescaling, eq. (\ref{stabeq1}) takes the form
\be 
-\ddot{\delta R}+\nabla^2 \delta R - (1-\frac{\bar{r}_c}{3\bar{r}} e^{-r})\delta R =0
\label{stabevol}
\ee
while for $m^2<0$ it can be written as
\be 
-\ddot{\delta R}+\nabla^2 \delta R + (1-\frac{\bar{r}_c}{3\bar{r}} \cos(r))\delta R =0
\label{instabevol}
\ee
where we have set the arbitrary phase $\theta$ to 0. In order to demonstrate the effects of the nonlinear term on the evolution of the perturbations we have considered the cases $\bar{r}_c=0$ (continuous lines on Fig. \ref{stability}) and $\bar{r}_c=10$ (dashed lines on Fig. \ref{stability}). The evolution of the perturbations is shown in Fig. \ref{stability} for $m^2>0$ (red lines) and $m^2<0$ (blue lines). Clearly, the evolution of the perturbations are significantly affected by the presence of the nonlinear term (dashed lines) but it appears that in this case, this term can not change the stability properties. In different forms of $f(R)$ the effects of such nonlinear terms may become important enough to stabilize such oscillating solutions or destabilize solutions that are stable at the linear level. Notice for example that one effect of the nonlinear term is the increase of the amplitude of the oscillations in the stable solution ($m^2>0$) thus leading to time-dependent oscillating terms which do not decay even though $m^2>0$. This is an interesting new effect with possible cosmological implications \cite{Steinhardt:1994vs-temporal-oscillations,Perivolaropoulos:2003we-temporal-oscillations} and deserves further investigation.

\subsection{Non-local Gravity Theories}

An important problem of GR is its behavior at small scales where it predicts the existence of singularities. In addition, at the quantum level the theory is plagued with unrenormalisable UV divergences (it is not UV finite)\cite{Goroff:1985th}. A possible cure of these divergences is the introduction of higher derivative terms in the Einstein-Hilbert action which can make the theory  UV finite \cite{Stelle:1977ry}. However, such terms introduce instabilities at the quantum level (a spin 2 component of the graviton propagator) which can also destabilise the classical vacuum of the theory. These instabilities can be cured by making the theory nonlocal through the introduction of infinite derivatives in the action leading to modification of the graviton propagator \cite{Biswas:2011ar}. In order to avoid the introduction of new poles, such infinite derivatives may be introduced in the form of an exponential of an entire function \cite{Tomboulis:1997gg,Siegel:2003vt,Deser:1986xr,Modesto:2011kw}.

\begin{figure}[!t]
\centering
\vspace{0cm}\rotatebox{0}{\vspace{0cm}\hspace{0cm}\resizebox{0.49\textwidth}{!}{\includegraphics{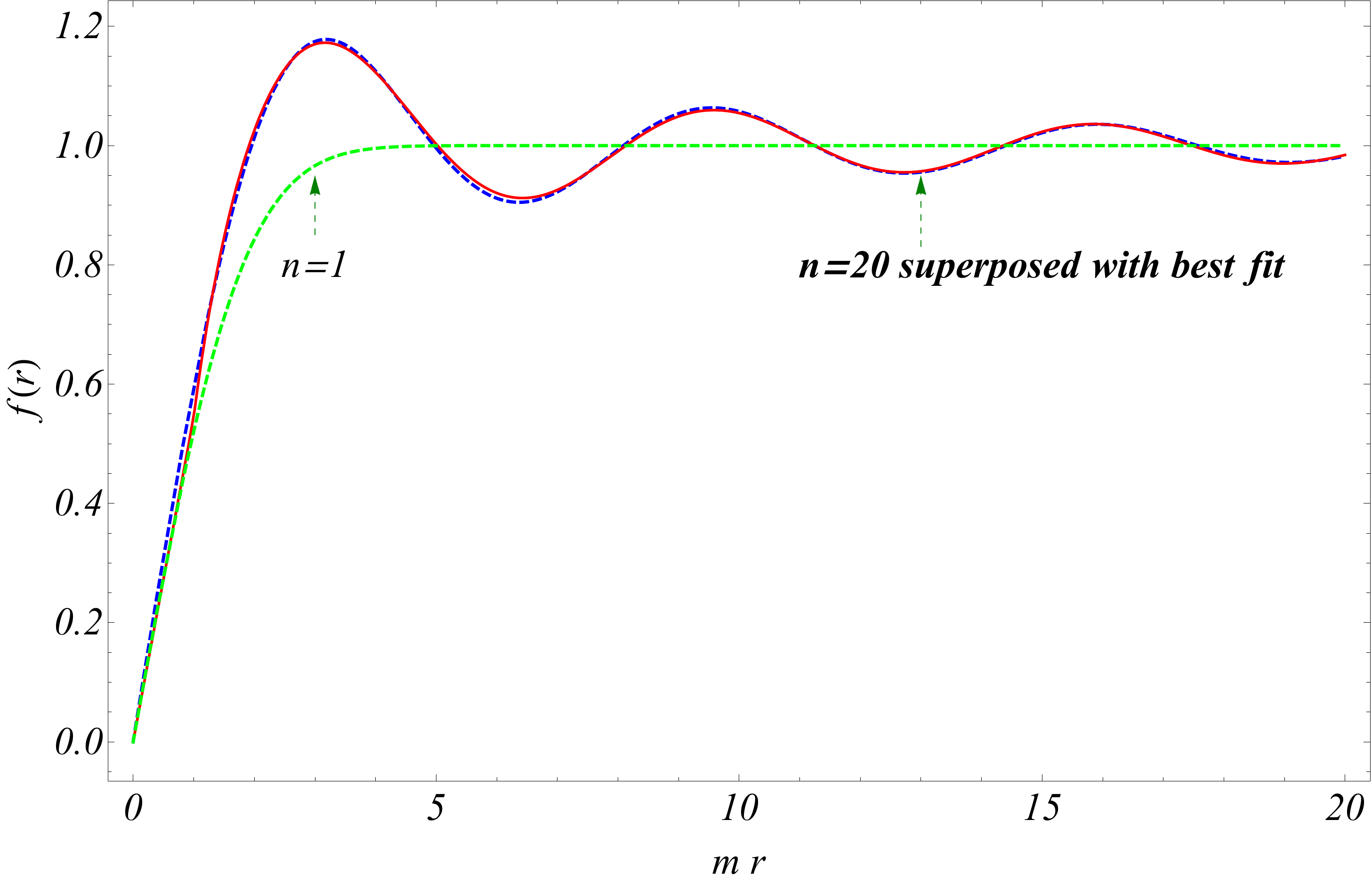}}}
\caption{The form of $f(r)$ for $n=1$ (dashed green line) and for $n=20$ superposed with the fit of (\ref{fitfr1}),(\ref{fitfr2}). For large $n$ there are decaying spatial oscillations for $\bar r \equiv m r \gsim 1$ which are very well fit by (\ref{fitfr1}),(\ref{fitfr2}). The linear behavior close to the origin dissolves the divergence of the Newtonian potential.}
\label{nonlocalfr}
\end{figure}

This class of nonlocal gravity theories generically softens UV divergences at the quantum level while removing the Big Bang and Black Hole singularities\cite{Biswas:2011ar,Frolov:2008uf-singularity-problem-forfR}. It also leads to a modification of the Newtonian potential around and below the scale of non-locality $m$\citep{Edholm:2016hbt-nonlocal-potential-stable-spatial-oscillations,Frolov:2015usa-newton-potential-nonlocal,Hohmann:2016yfd-spatial-oscil-non-local,Kehagias:2014sda-nonlocal-oscillations,Maggiore:2014sia-nonlocal-gravity-oscil}. This modification includes a removal of the divergence of the potential at $r=0$ (the potential goes to a constant at $r=0$) and the possibility of the introduction of decaying spatial oscillations on scales close to the scale of non-locality. In particular, the predicted form of the modified Newtonian potential in these theories is of the form \cite{Edholm:2016hbt-nonlocal-potential-stable-spatial-oscillations}
\be 
V_{eff}(r)= -\frac{G M}{r} f(r,m)
\label{newtpotnonlocal}
\ee
where
\be 
f(r,m)=\frac{1}{\pi} \int_{-\infty}^{+\infty} dk \frac{sin(k r) e^{-\tau(k,m)}}{k}
\label{frm-lonloc}
\ee
A typical form for  $\tau$ is 
\be
\tau=\frac{k^{2n}}{m^{2n}}
\label{tauform}
\ee
For $n=1$ it may be shown that $f(r)=Erf(m\frac{r}{2})$ which is linear $\sim r$ for $r<m^{-1}$ and goes to a constant for $r\gg m^{-1}$. The form of $f(r)$ for $n=1$ and $n=20$ is shown in Fig. \ref{nonlocalfr} ($\bar r \equiv m r$). For $n>10$ the form of $f(r)$ is practically unchanged. For large $n$, $f(r)$ is very well fit by the function
\ba  
f(r)&=&\alpha_1 \bar r \hspace{0.2cm} 0<\bar r<1 
\label{fitfr1}\\
f(r)&=&1 + \alpha_2 \frac{\cos(\bar r+\theta)}{\bar r} \hspace{0.2cm} 1<\bar r
\label{fitfr2}
\ea
where $\alpha_1=0.544$, $\alpha_2 = 0.572$, $\theta=0.885 \pi$.
Notice the decaying oscillations that develop for $r\gsim m^{-1}$ which constitute a signature for this class of models. This class of models are particularly interesting not only for their UV finiteness but also because they are free from singularities while having a well defined Newtonian limit in the case $n=1$. It is therefore important search for this type of spatially oscillating signature in torsion balance experiment data. This type of test is implemented in the next section.

\section{Oscillating corrections on Newton's constant: Consistency with macroscopic observations and torsion balance experiments}
\label{sec:Section 3}
\subsection{Observational viability of spatial oscillations of Newton constant}

\begin{figure}[!t]
\centering
\vspace{0cm}\rotatebox{0}{\vspace{0cm}\hspace{0cm}\resizebox{0.49\textwidth}{!}{\includegraphics{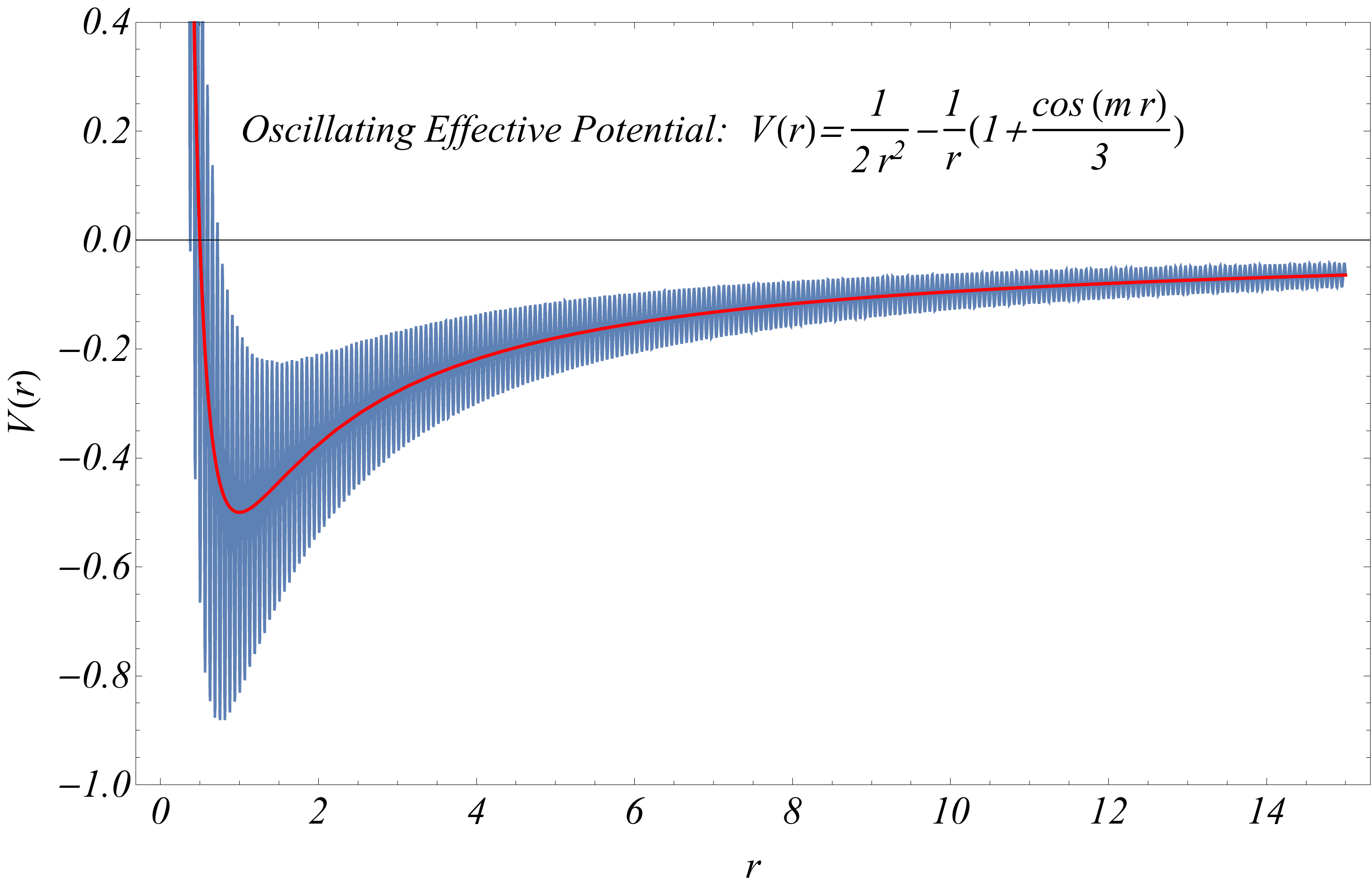}}}
\caption{The oscillating effective potential for a bound system of eq. (\ref{boundpot}) with $m=100$, superposed with the corresponding Newtonian potential (red line).}
\label{boundpot}
\end{figure}

The generalized Newtonian force forms predicted in the weak field limit of the theoretical models discussed in the previous section may be obtained easily by differentiation of the corresponding effective gravitational potentials.
For the non-local large $n$ effective potential (\ref{newtpotnonlocal}) fit by eqs. (\ref{fitfr1}), (\ref{fitfr2})  we obtain for $r>m^{-1}$
\be 
\vec{F}_1=-\hat{r}\frac{G M}{r^2}\left(1+\frac{2\alpha_2 \cos(m r +\theta)}{m r}+\alpha_2 \sin(m r +\theta)\right)
\label{forcenonlocal}
\ee
while for the oscillating effective potential (\ref{ntpot2}) obtained for $f(R)$ theories we have
\be 
\vec{F}_2=-\hat{r}\frac{G M}{r^2}\left(1+\frac{\cos(m r +\theta)}{3}+\frac{m r}{3} \sin(m r +\theta)\right)
\label{forcefR}
\ee
The radial weak field geodesic equation for a bound system in such a force field, after proper rescaling is of the form
\be 
\ddot{r}=\frac{1}{r^3}-\frac{1}{r^2}\left(1+\frac{1}{3}\cos(m r +\theta)+\frac{1}{3} m r \sin(m r +\theta)\right)
\label{radgeod}
\ee
This equation may be obtained using the effective potential
\be 
V(r)=\frac{1}{2 r^2} - \frac{1}{r}\left(1+\frac{1}{3} \cos(m r +\theta)\right)
\label{voscbound}
\ee
which is shown in Fig. \ref{boundpot} along with the corresponding Newtonian effective potential of a bound system.

Both types of forces (\ref{forcenonlocal}) and (\ref{forcefR})  clearly do not have a Newtonian limit since as $m\rightarrow \infty$ there is no well defined limit for the corresponding deviations. However, the existence of a Newtonian limit is not the relevant question for the viability of these predictions. The relevant question is the following: `What are the experimental and observational consequences of the above forms of gravitational forces  for
\be
m> 10 mm^{-1}
\label{mrange}
\ee
and are these consistent with current experiments and observations?' 

In view of the fact that the extra spatially oscillating force component averages out to 0 over scales larger than $1mm$ makes the answer to this question a nontrivial issue. Immediately ruling out these oscillating gravitational force forms just because they do not have a Newtonian limit would be like ruling out quantum theory because it predicts that macroscopic objects have a wave nature before checking if their de Broglie wavelength is consistent with experiments.

\begin{figure}[!t]
\centering
\vspace{0cm}\rotatebox{0}{\vspace{0cm}\hspace{0cm}\resizebox{0.49\textwidth}{!}{\includegraphics{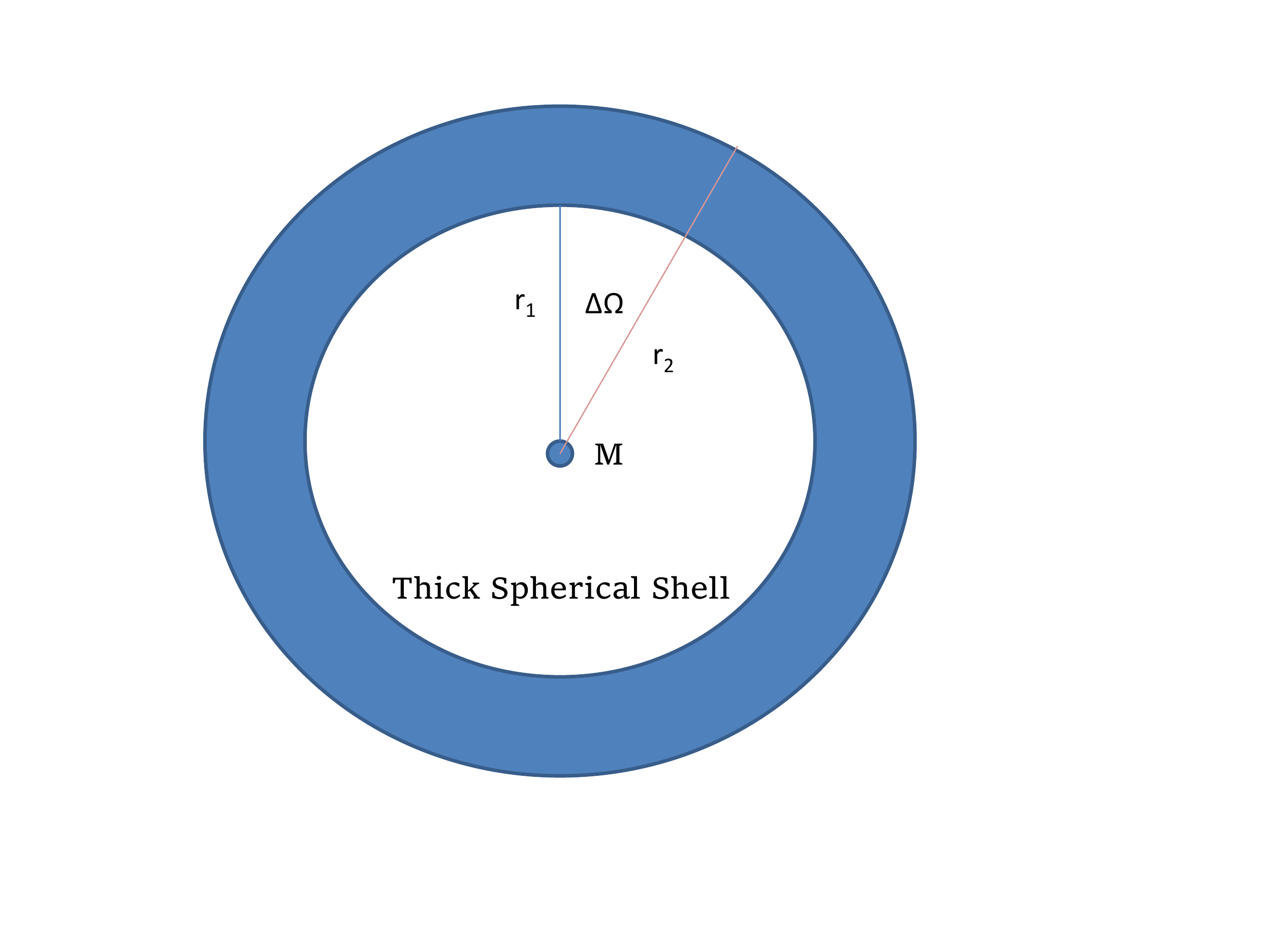}}}
\caption{A point mass interacting gravitationally with a thick massive spherical shell.}
\label{spher-shell}
\end{figure}

As a simple useful toy model-system where the detectability of the predicted force oscillations can be tested consider a point mass at the center of a thick homogeneous spherical shell of density $\rho_{sh}$ with inner radius $r_1$ and outer radius $r_2$ (Fig. \ref{spher-shell}). Consider now the magnitude of the force exerted by a small solid angle $\Delta \Omega$ of the shell. Using eq. (\ref{forcefR}) (where the oscillating terms are more important compared to the corresponding terms in eq. (\ref{forcenonlocal})) we can easily find the force magnitude per solid angle exerted by the shell on the particle as
\be 
F_{tot}=F_N+F_a+F_b
\label{forceshell1}
\ee
where $F_N$ is the Newtonian contribution and $F_a$, $F_b$ are the contributions from the oscillating components which are of the form
\ba 
F_N(r_1,r_2)&=&G M \rho_{sh} (r_2-r_1)\\
F_a(r_1,r_2)&=&G M \rho_{sh} \frac{2}{m}(\sin(m r_2)-\sin(m r_1))\\
F_b(r_1,r_2)&=& G M \rho_{sh} (r_2 \cos(m r_2)- r_1 \cos(m r_1))
\label{forcecompb}
\ea
For large $r_i$ and thin shell we can have $F_b >> F_N$ and thus it would appear that the oscillating contribution contradicts all current experiments and observations. However, what is actually measured in any experiment or observation is an {\it average} gravitational force over a range of distances and object dimensions. This averaging is due to the relative motion of objects during an observation and also due to the inaccurate knowledge of distances and dimensions of gravitating objects or even due to the quantum uncertainty principle. Thus, what the observable gravitational force is
\be 
\bar{F}=\frac{1}{\delta r^2} \int_{r_1-\delta r/2}^{r_1+\delta r/2} \int_{r_2-\delta r/2}^{r_2+\delta r/2} dr'_1 dr'_2 \;F(r'_1,r'_2)
\label{fobsaverage}
\ee
where $\delta r$ is the uncertainty in $r_1$, $r_2$ due to the above mentioned factors. It is straightforward to show that for any finite $r_1$, $r_2$, $\delta r$ and large enough $m$ the ratios of the oscillating components over the Newtonian component obey 
\ba
\frac{\bar F_b}{\bar F_N}& \lsim & O\left((m \delta r)^{-1}\right) \\
\frac{\bar F_a}{\bar F_N}  &\lsim & O\left((m \delta r)^{-1}\right) 
\label{fafnbounds}
\ea
and thus can be made arbitrarily small and consistent with macroscopic observations and experiments. A similar conclusion can be obtained for the oscillating components of eq. (\ref{forcenonlocal}) for the oscillations coming from the nonlocal gravity theory.

In view of the derived consistency of the predicted oscillating gravitational force terms with macroscopic observations it is clear that oscillating force signatures of these theories can be searched in laboratory experiments testing the validity of the Newtonian potential at sub-millimeter scales. The most constraining such experiment to date for the particular types of potentials discussed in the present analysis is the torsion balance experiments of the Washington group\cite{Kapner:2006si-washington3}. The Newtonian residual torque data obtained by the Washington group have been used to constrain Yukawa and power law type corrections to the Newtonian potential of the form 
\be V_{eff}=-\frac{G M}{r}\left(1+\alpha e^{-m r}\right) \label{ntpot-wash} \ee
and have ruled out values of $m\lsim 18mm^{-1}$ for $\alpha=1$ at the $2\sigma$ level. 
The corresponding gravitational force oscillating parametrization is of the form
\be F_{eff}=-\hat{r}\frac{G M}{r^2}\left(1+\alpha \; e^{-m r}+\alpha m r\; e^{-m r}\right)
\label{ntforceexp1}
\ee 

\begin{figure*}[ht]
\centering
\begin{center}
$\begin{array}{@{\hspace{-0.10in}}c@{\hspace{0.0in}}c}
\multicolumn{1}{l}{\mbox{}} &
\multicolumn{1}{l}{\mbox{}} \\ [-0.2in]
\epsfxsize=3.3in
\epsffile{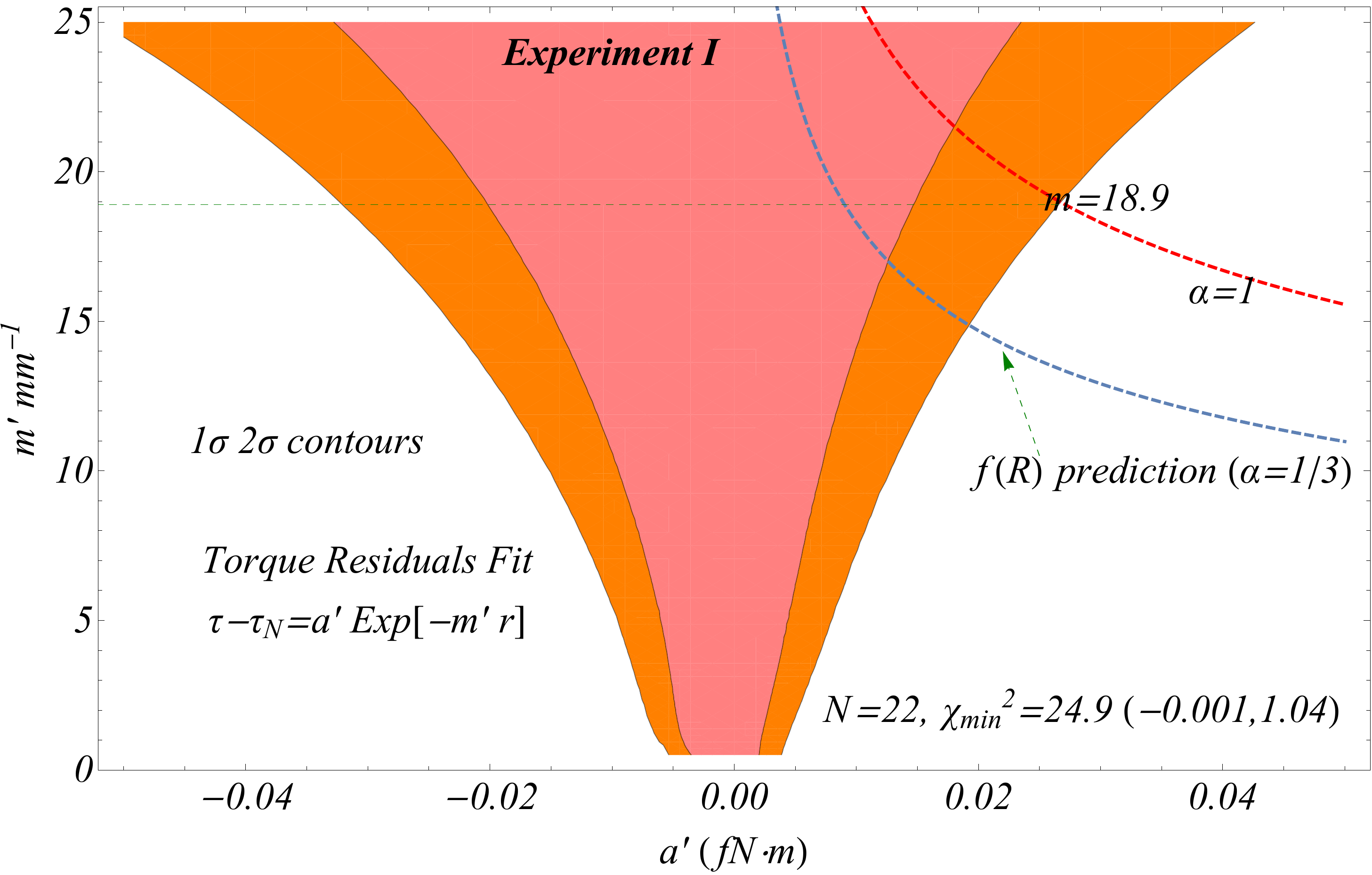} &
\epsfxsize=3.3in
\epsffile{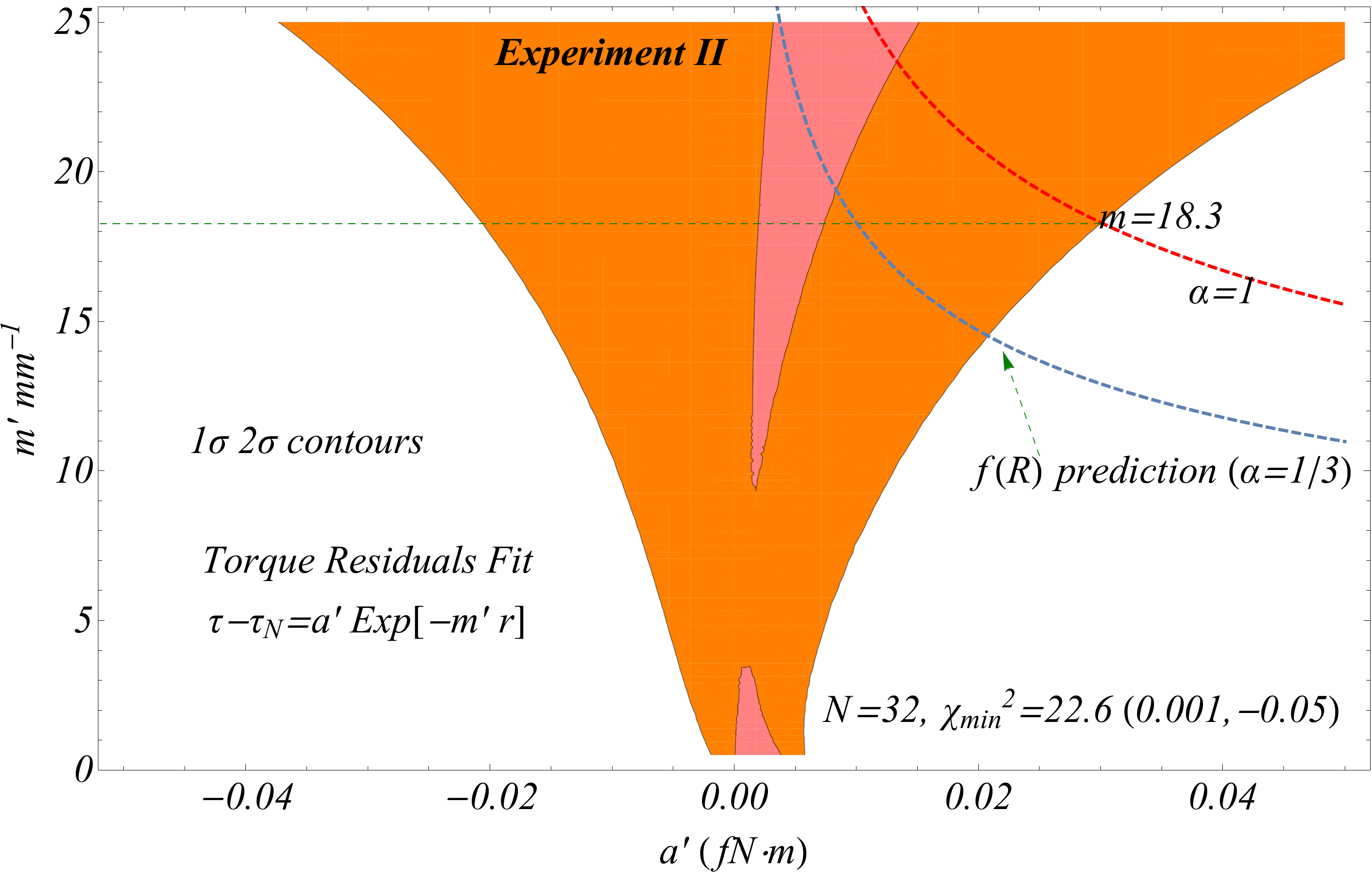} \\
\epsfxsize=3.3in
\epsffile{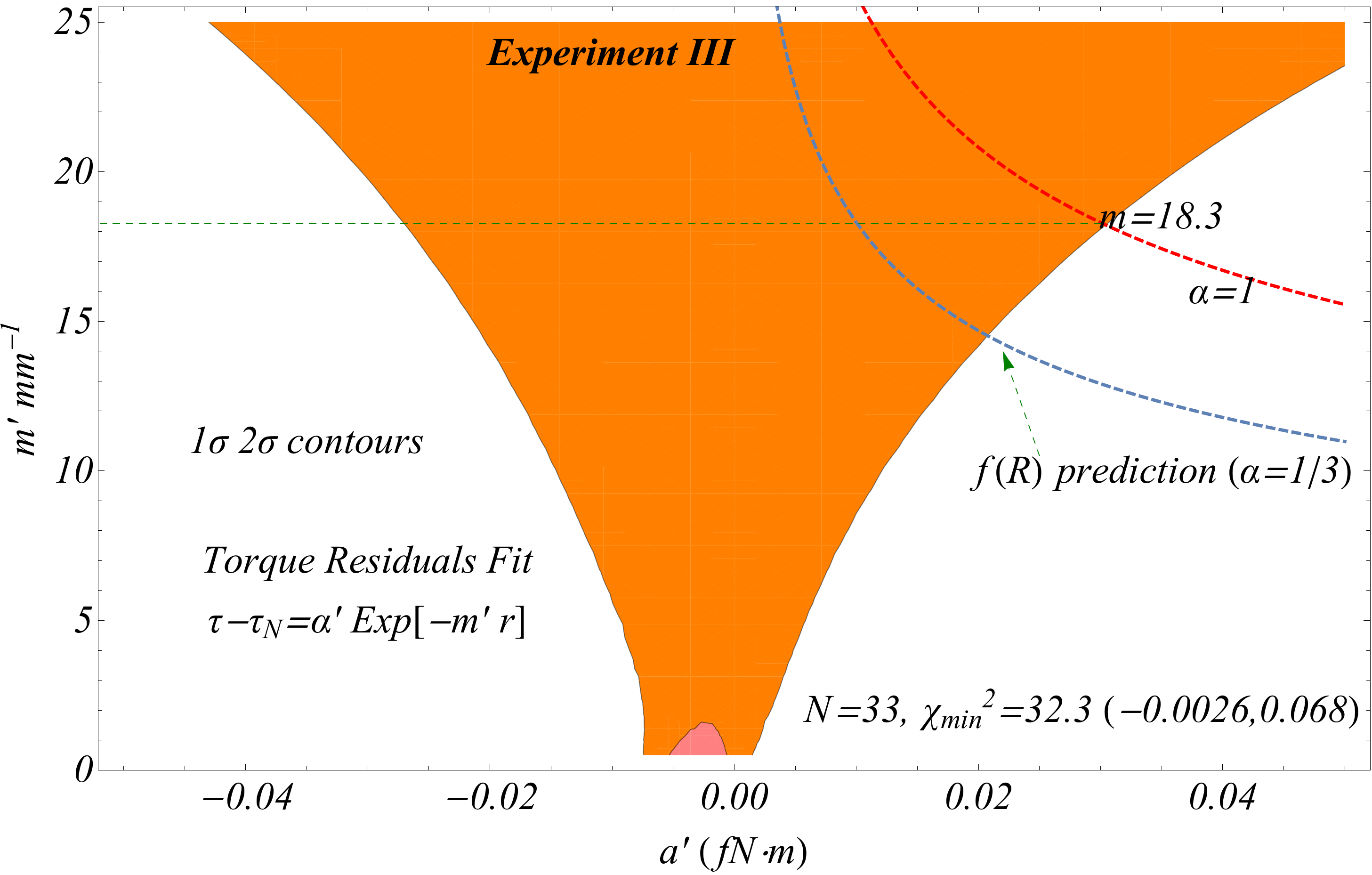} &
\epsfxsize=3.3in
\epsffile{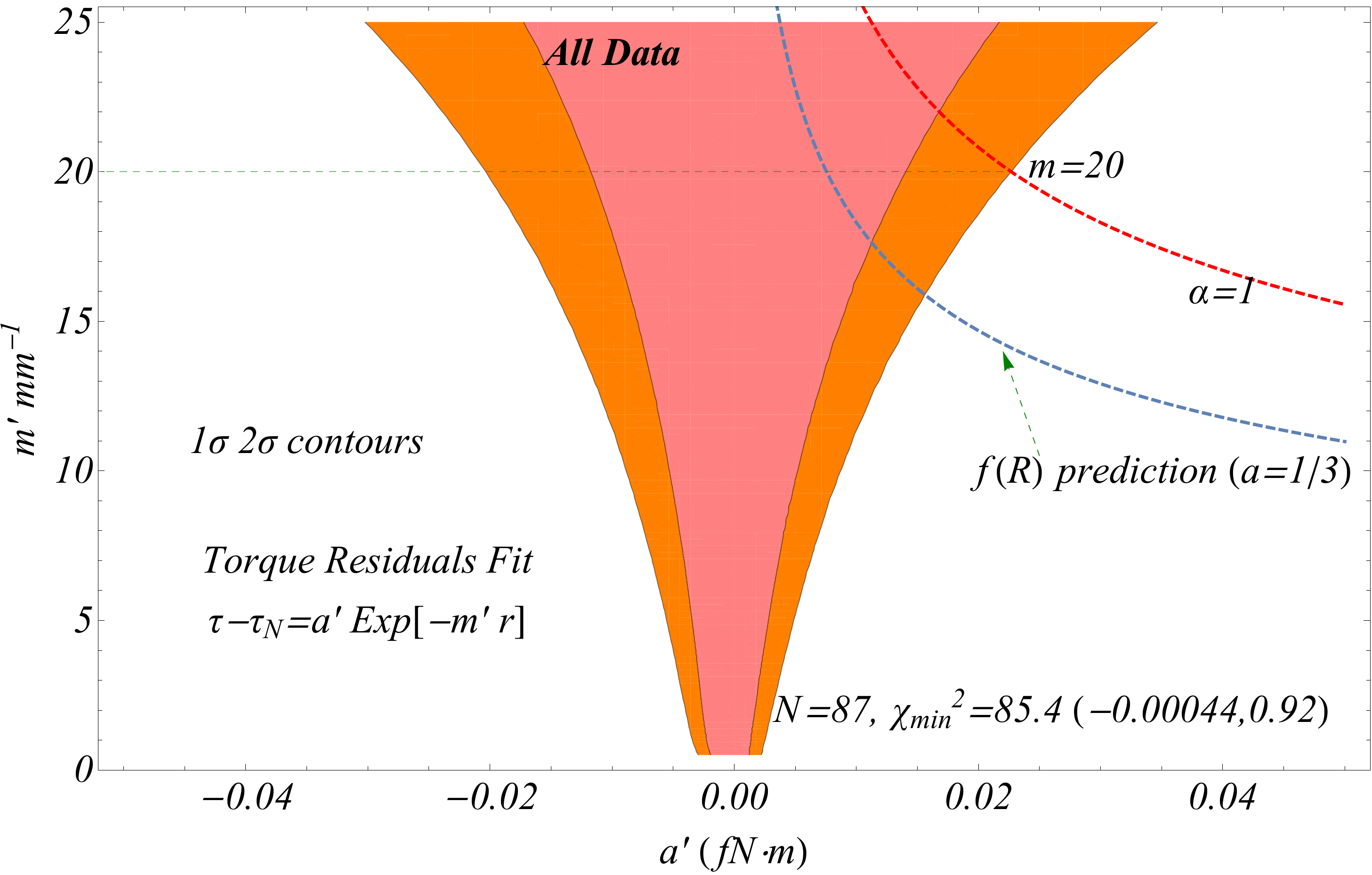}\\
\end{array}$
\end{center}
\vspace{0.0cm}
\caption{\small The $1\sigma$ and $2\sigma$ contours in the parameter space $(\alpha',m')$  for the Yukawa parametrization. The blue dashed line is the line $\alpha=\frac{1}{3}$ projected onto the parameter space $(\alpha',m')$ in accordance with the empirical relations (\ref{mcon}), (\ref{alphcon}). The red line corresponds to $\alpha=1$ obtained from (\ref{mcon}), (\ref{alphcon}) and intersects the $2\sigma$ contour at $m\simeq 20 mm^{-1}$ leading to a $2\sigma$ constraint $m>20 mm^{-1}$ in good agreement with the published $2\sigma$ constraint on $m$ by the Washington group\cite{Kapner:2006si-washington3}. This agreement is a good test for the validity of our analysis. }
\label{figcontoursexp}
\end{figure*}

In the next section we extend the analysis of the Washington group \citep{Kapner:2006si-washington3}  using a oscillating parametrization in addition to the Yukawa one in an attempt to search for oscillating signatures or constraints in the data. We test the validity of our analysis by veryfing that we obtain the same bound on $m$ as the one of Ref. \cite{Kapner:2006si-washington3} for the Yukawa parametrization.

\subsection{Fitting spatial oscillations of Newton's constant to the Washington experiment data}

\subsubsection{Maximum Likelihood Analysis}

The Washington experiment used a missing mass torsion balance instrument measuring gravitational interaction torques with extreme accuracy. The torques developed between missing masses (holes) present in a torsion pendulum detector and similar holes present in a rotating with constant angular velocity attractor ring. 

The differences (residuals) between the measured torques and their expected Newtonian values were recorded in three experiments (I, II, III) using the same detector but different thickness of attractor disks. The attractor thickness in each experiment was chosen in such a way as to reduce systematic errors by comparing the residuals among the three experiments. The residual torques for each experiment as a function of detector-attractor separation were published in three Figures (one for each experiment). 
Experiment I suffered a minor systematic effect (the detector ring was found to be slightly bowed) which was accounted for by modeling the heights of the outer sets of holes to different heights. No such systematic was present in Experiments II and III.

A total of 87 residual points were shown along with three predicted residual curves \cite{Kapner:2006si-washington3,Hoyle:2000cv-washington1,Hoyle:2004cw-washington2,Kapner:2005qy-thesis}  that would arise in the context of Yukawa type deviations (eq. (\ref{ntpot-wash})) from the Newtonian potential for three pairs of $(\alpha,\lambda\equiv 1/m)$.
Each point referred to the value of the residual torque (measured value minus Newtonian prediction), the attractor-detector distance in $mm$ and the $1\sigma$ error of the residual torque (see Appendix Table II). 

\begin{figure*}[ht]
\centering
\begin{center}
$\begin{array}{@{\hspace{-0.10in}}c@{\hspace{0.0in}}c}
\multicolumn{1}{l}{\mbox{}} &
\multicolumn{1}{l}{\mbox{}} \\ [-0.2in]
\epsfxsize=3.3in
\epsffile{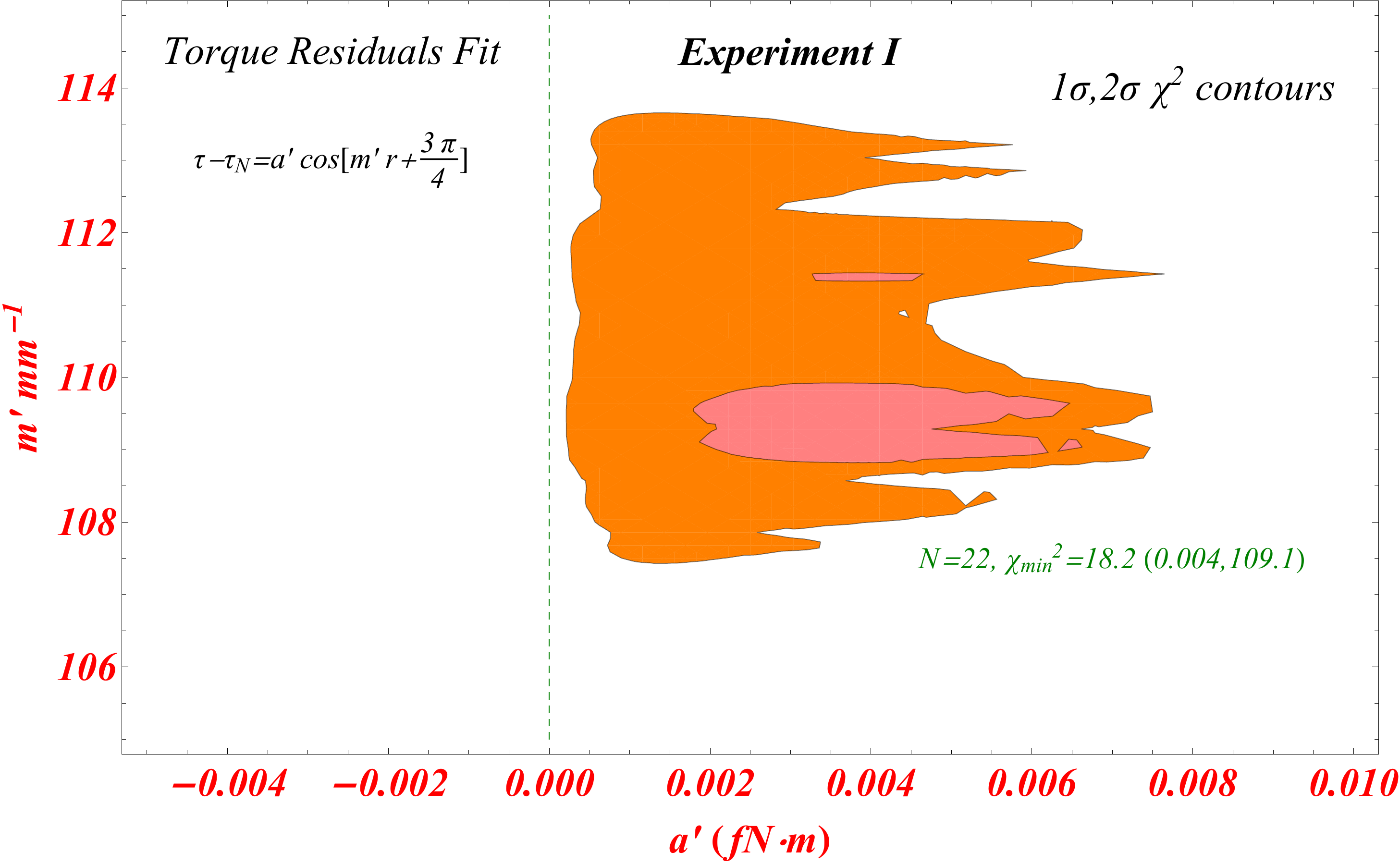} &
\epsfxsize=3.3in
\epsffile{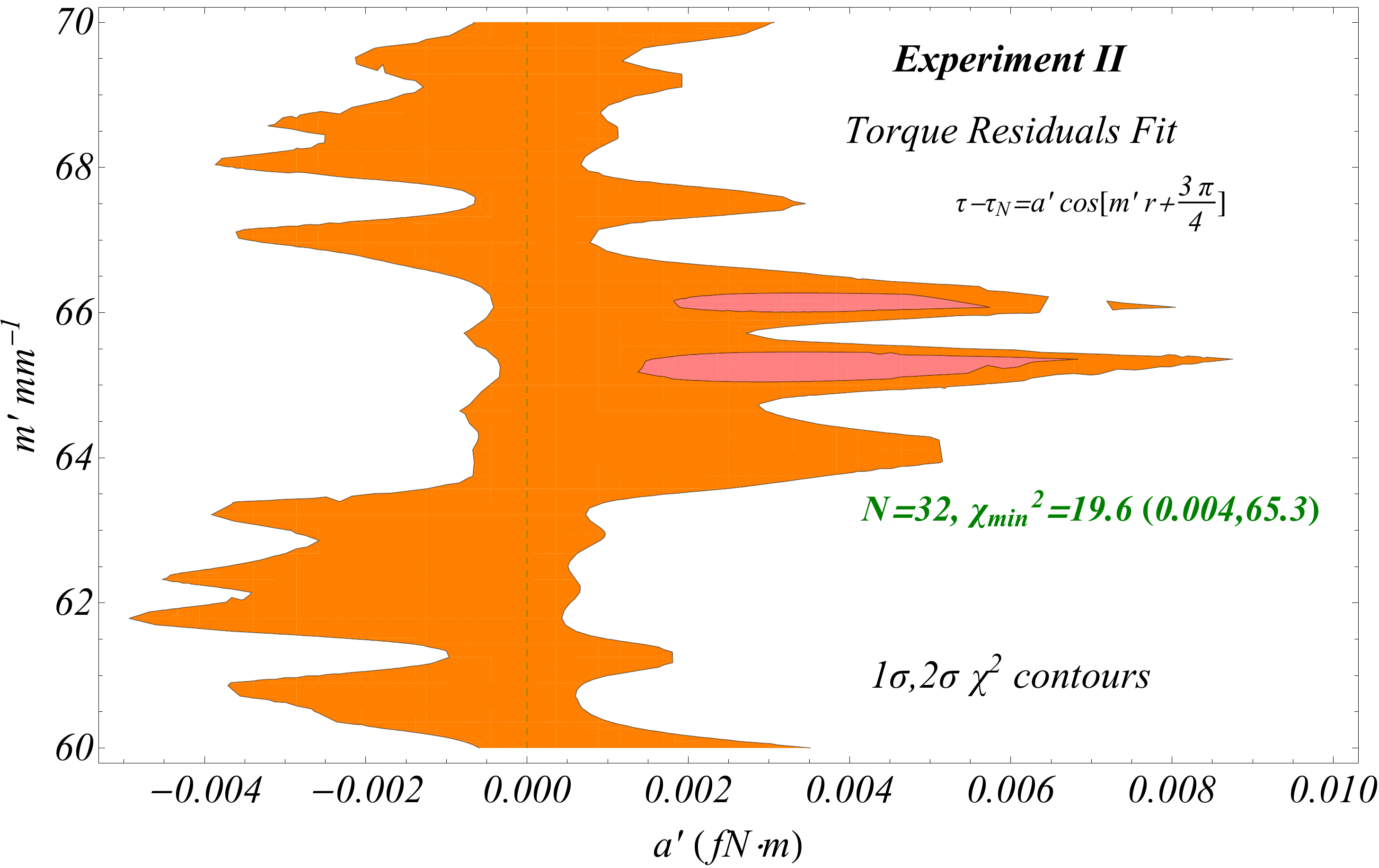} \\
\epsfxsize=3.3in
\epsffile{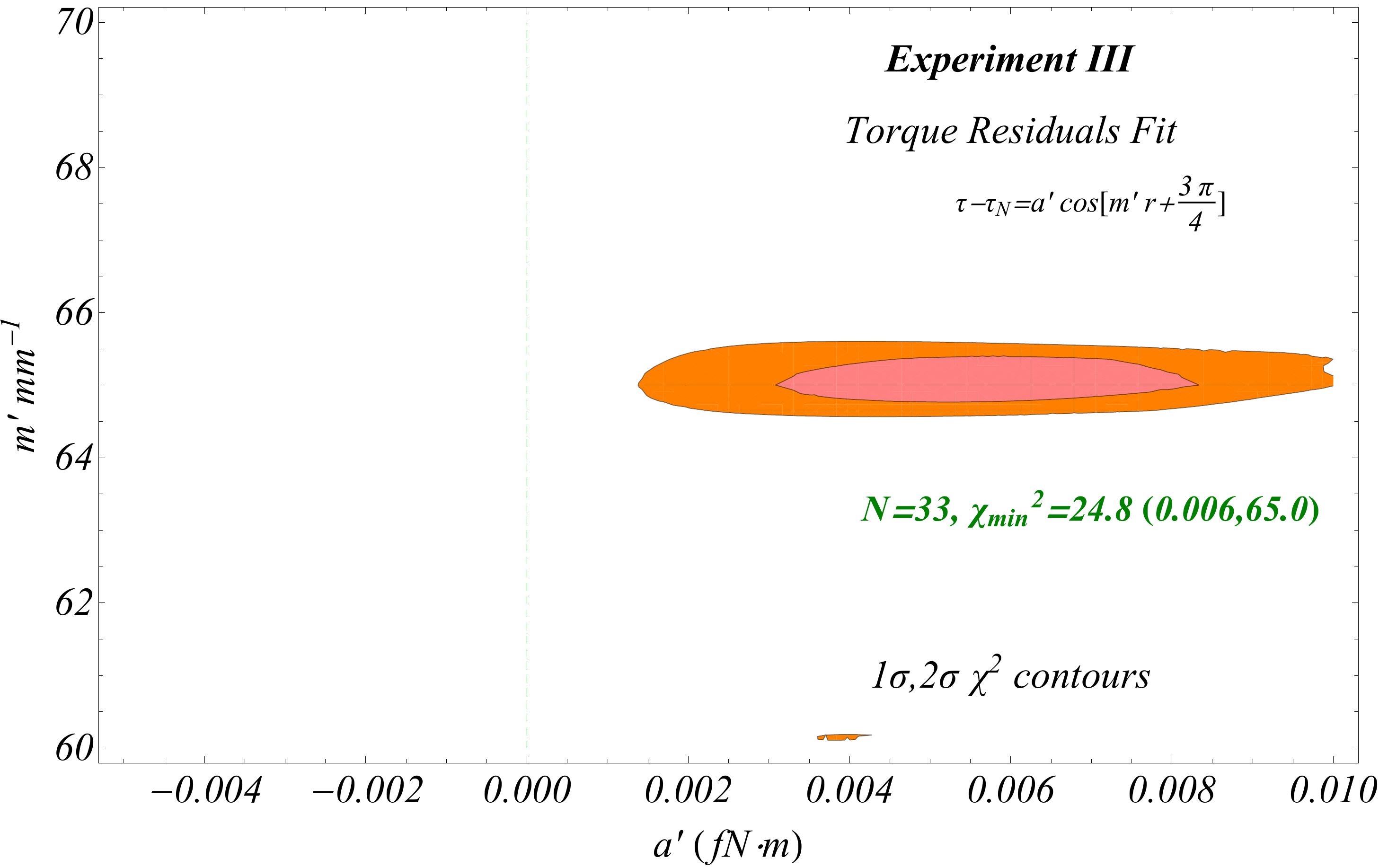} &
\epsfxsize=3.3in
\epsffile{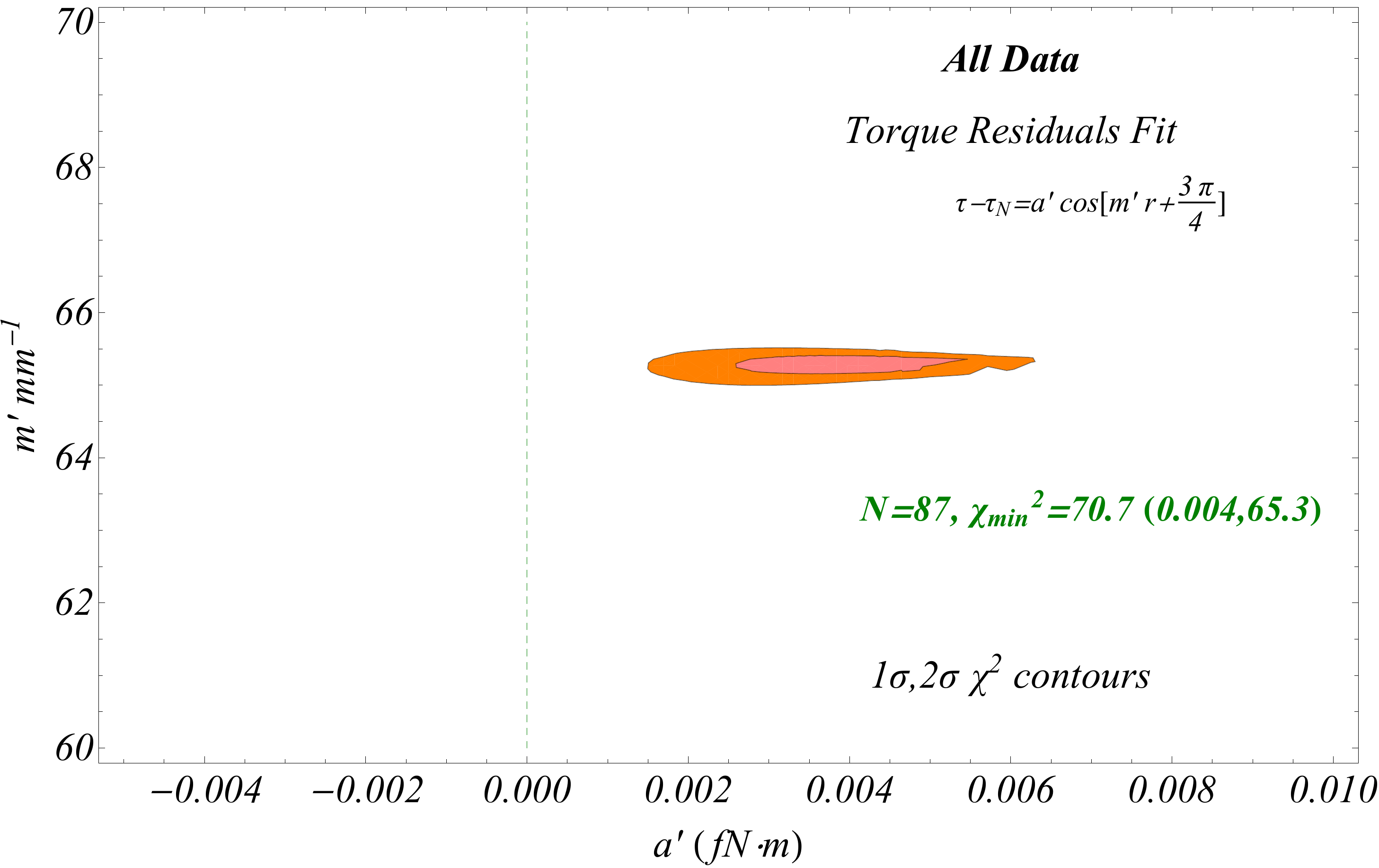}\\
\end{array}$
\end{center}
\vspace{0.0cm}
\caption{\small The $1\sigma$ and $2\sigma$ contours in the parameter space $(\alpha',m')$  for the oscillating parametrization with $\theta'=\frac{3\pi}{4}$. Notice that experiment III appears to have the highest constraining power with respect to the oscillating parametrization while in Experiment I the spatial oscillations are best fit by a higher spatial frequency (we used red color for the framelabel of Experiment I to show this distinct behavior). However, even the data of Experiment I are well fit by the same spatial frequency as the other two experiments (local minimum of $\chi^2$ as shown in Fig. \ref{figchi2vsm}). For the combined dataset there is a well defined high quality fit at $(\alpha',m')=(0.004,65.3)$ corresponding to a wavelength $\lambda=\frac{2\pi}{m}=0.096mm$. This best fit is about $3\sigma$ away from the null Newtonian value $\alpha'=0$.}
\label{contoursoscil}
\end{figure*}

\begin{figure*}[ht]
\centering
\begin{center}
$\begin{array}{@{\hspace{-0.10in}}c@{\hspace{0.0in}}c}
\multicolumn{1}{l}{\mbox{}} &
\multicolumn{1}{l}{\mbox{}} \\ [-0.2in]
\epsfxsize=3.3in
\epsffile{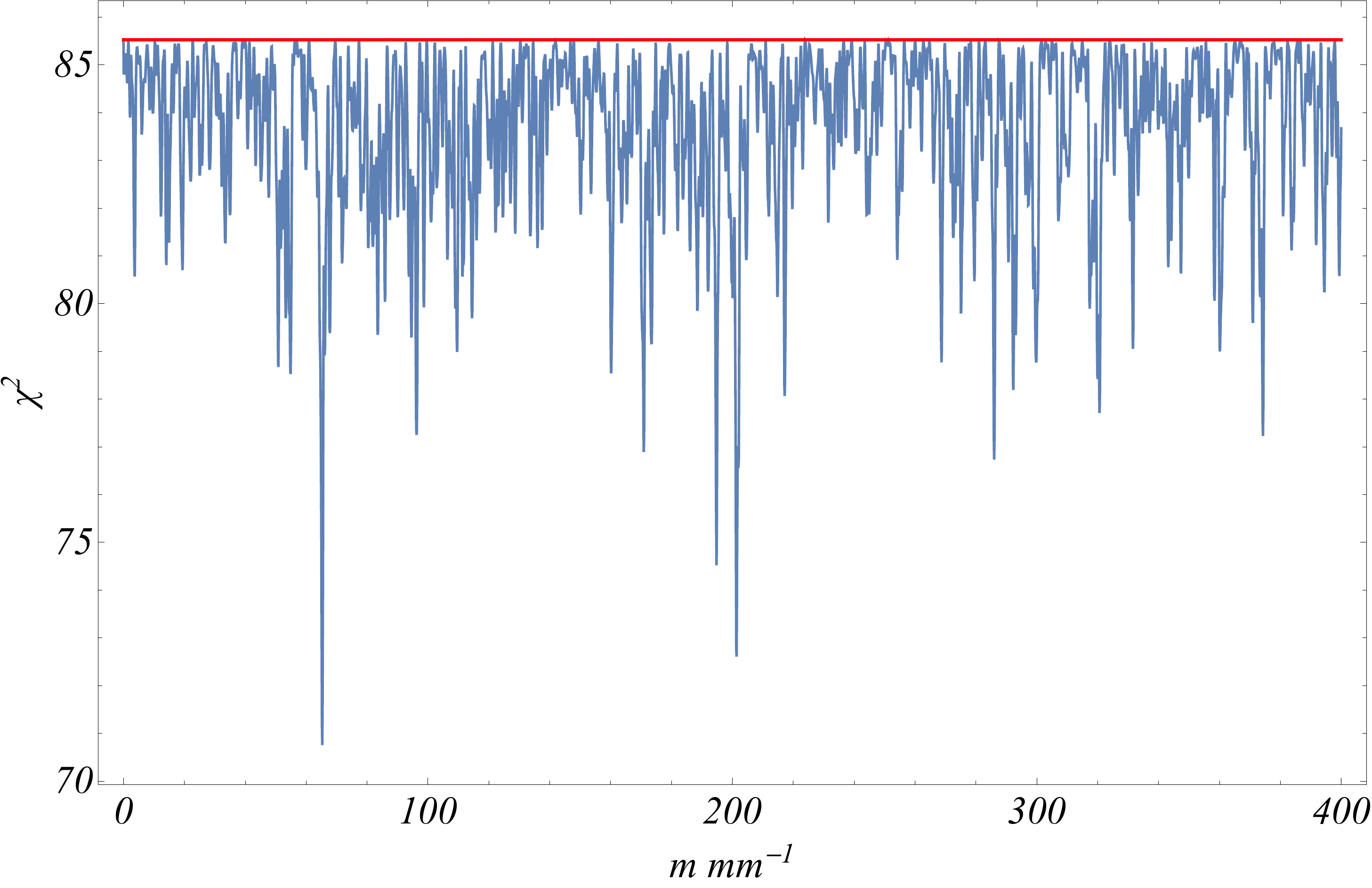} &
\epsfxsize=3.3in
\epsffile{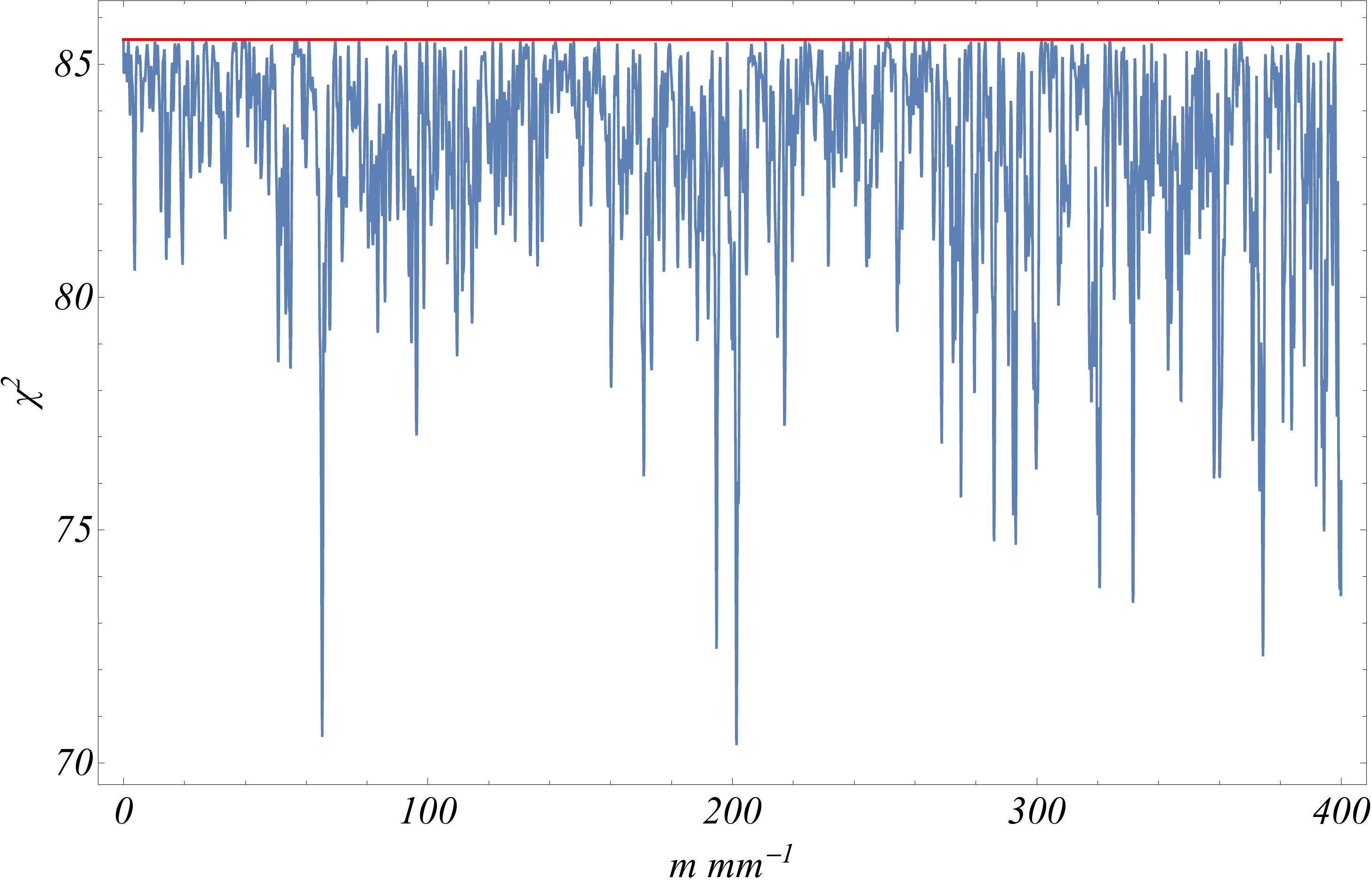} \\
\end{array}$
\end{center}
\vspace{0.0cm}
\caption{\small The value of the minimized $\chi^2$ as a function of the spatial frequency $m$ without including horizontal uncertainties (left panel) and including horizontal uncertainties $\sigma_r=0.002mm$ at each data-point. The location of the minima is not affected while their depth increases at higher spatial frequencies. The red straight line corresponds to the Newtonian residual $\delta \tau=0$.}
\label{figchi2vsm}
\end{figure*}

\begin{figure}[!t]
\centering
\vspace{0cm}\rotatebox{0}{\vspace{0cm}\hspace{0cm}\resizebox{0.49\textwidth}{!}{\includegraphics{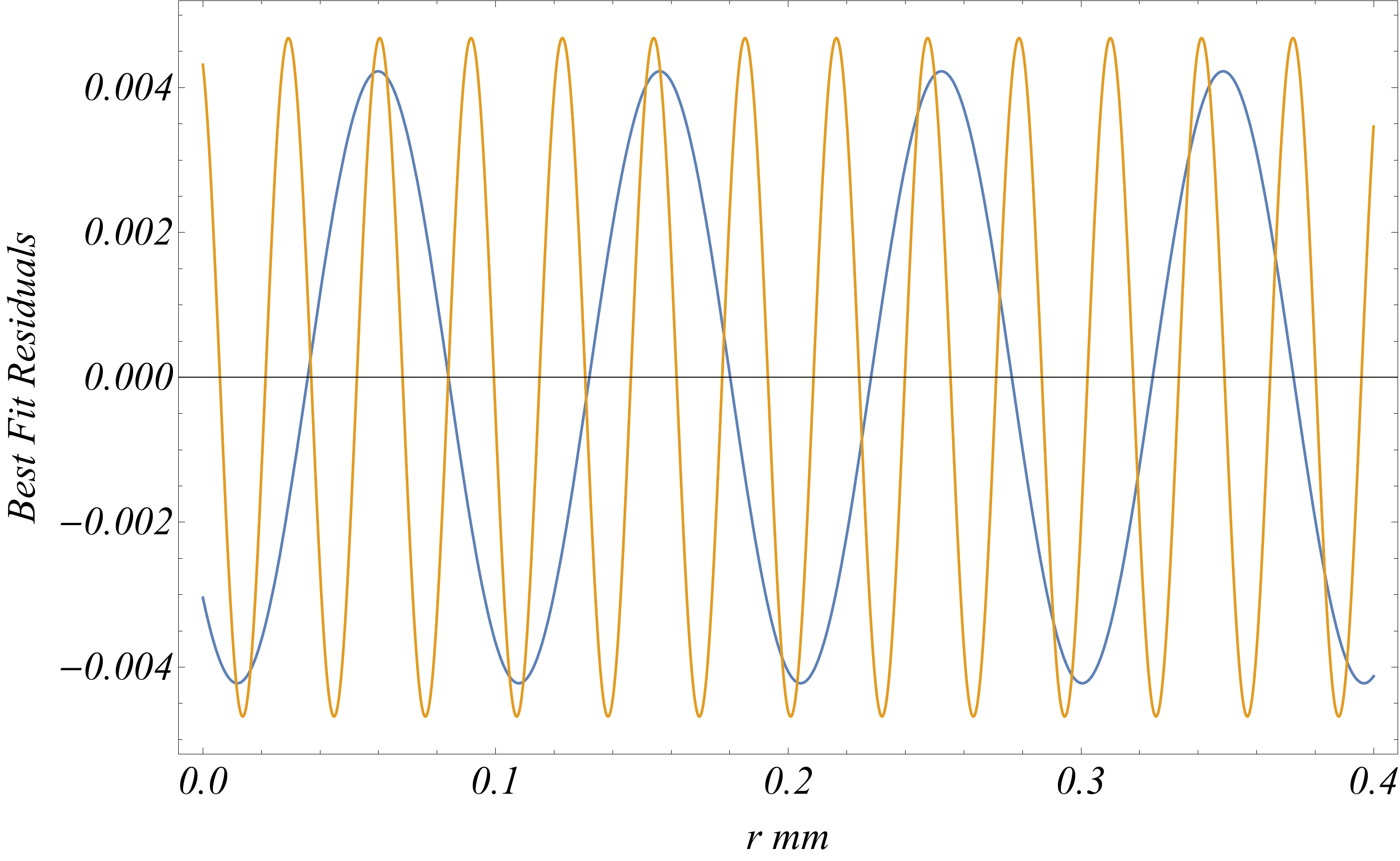}}}
\caption{The best fit oscillating forms corresponding to the two deepest $\chi^2$ minima ($m\simeq 65 mm^{-1}$ and $m\simeq 202mm^{-1}$). The blue curve with the larger wavelength, corresponds to a spatial frequency $m=65 mm^{-1}$ with wavelength $\lambda=2\pi/m\simeq 0.097mm=97\mu m$. The (approximate) higher harmonic (brown curve) corresponds to a spatial frequency $m=202 mm^{-1}$ with wavelength $\lambda=2\pi/m\simeq 0.031mm$. The roots and maxima of the two curves are correlated and the second curve (brown) appears as a higher harmonic of the first (blue).}
\label{bfitparams}
\end{figure}

\begin{figure*}[ht]
\centering
\vspace{1.0cm}
\begin{center}
$\begin{array}{@{\hspace{-0.10in}}c@{\hspace{0.0in}}c}
\multicolumn{1}{l}{\mbox{}} &
\multicolumn{1}{l}{\mbox{}} \\ [-0.2in]
\epsfxsize=3.3in
\epsffile{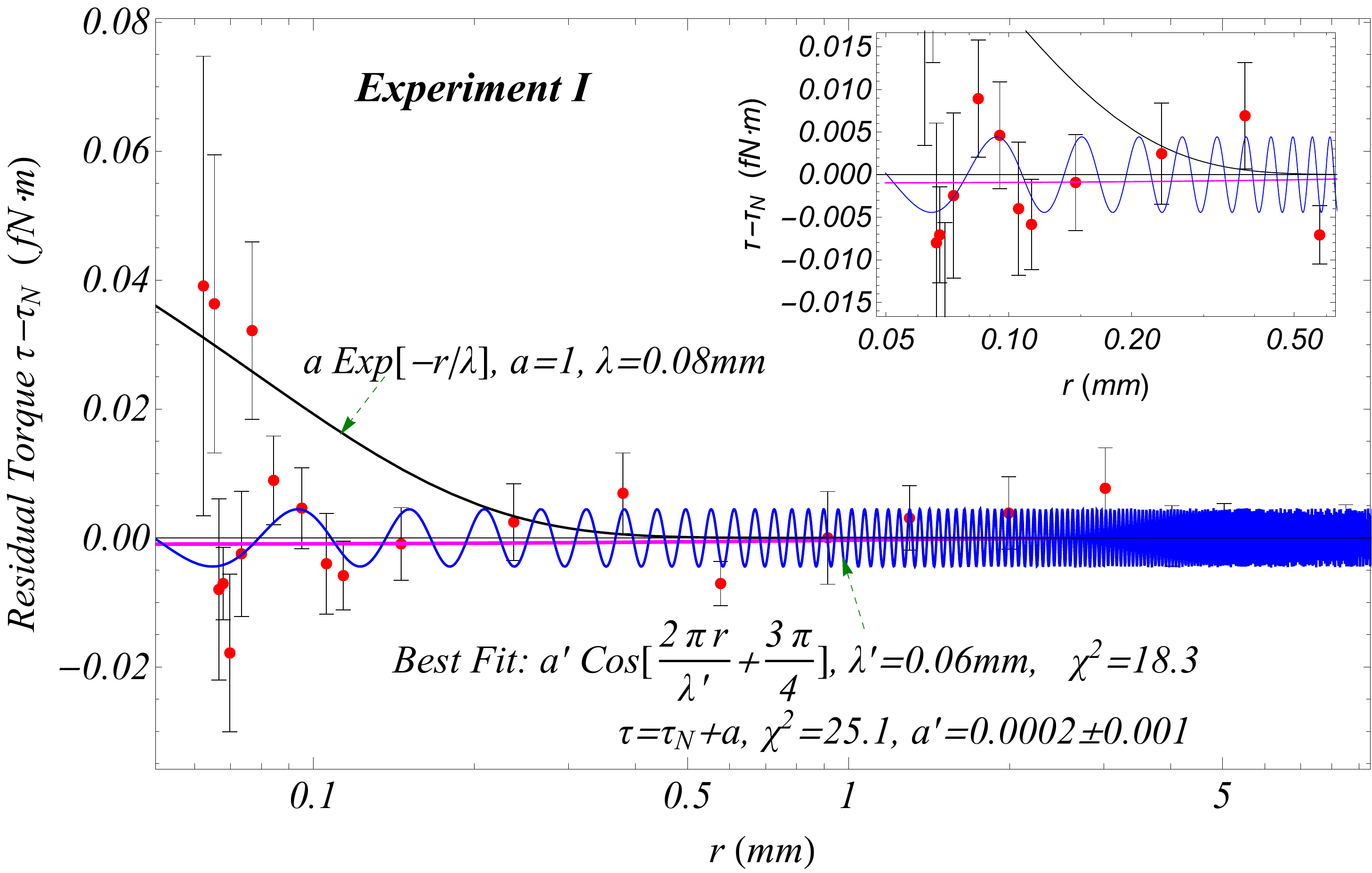} &
\epsfxsize=3.3in
\epsffile{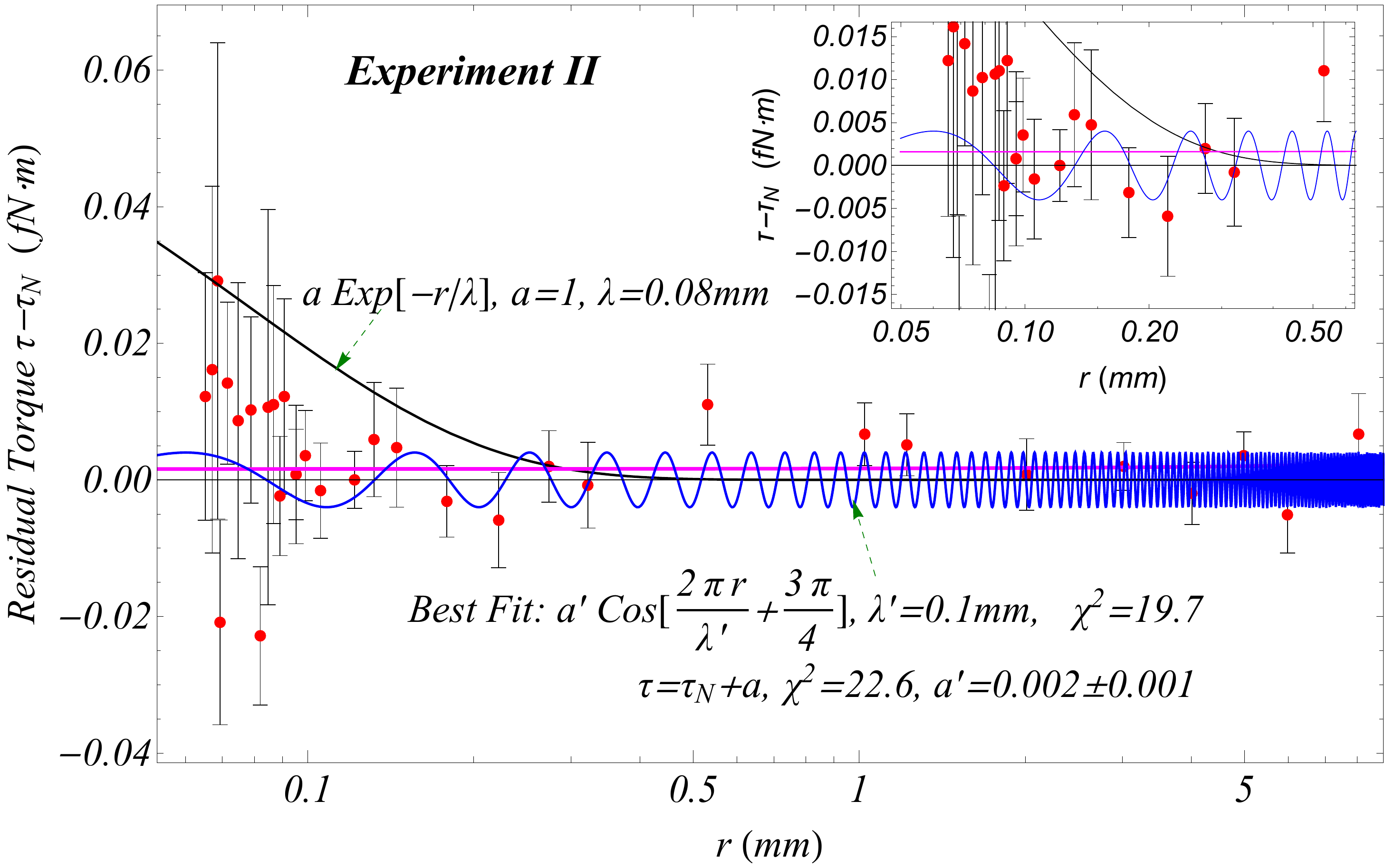} \\
\epsfxsize=3.3in
\epsffile{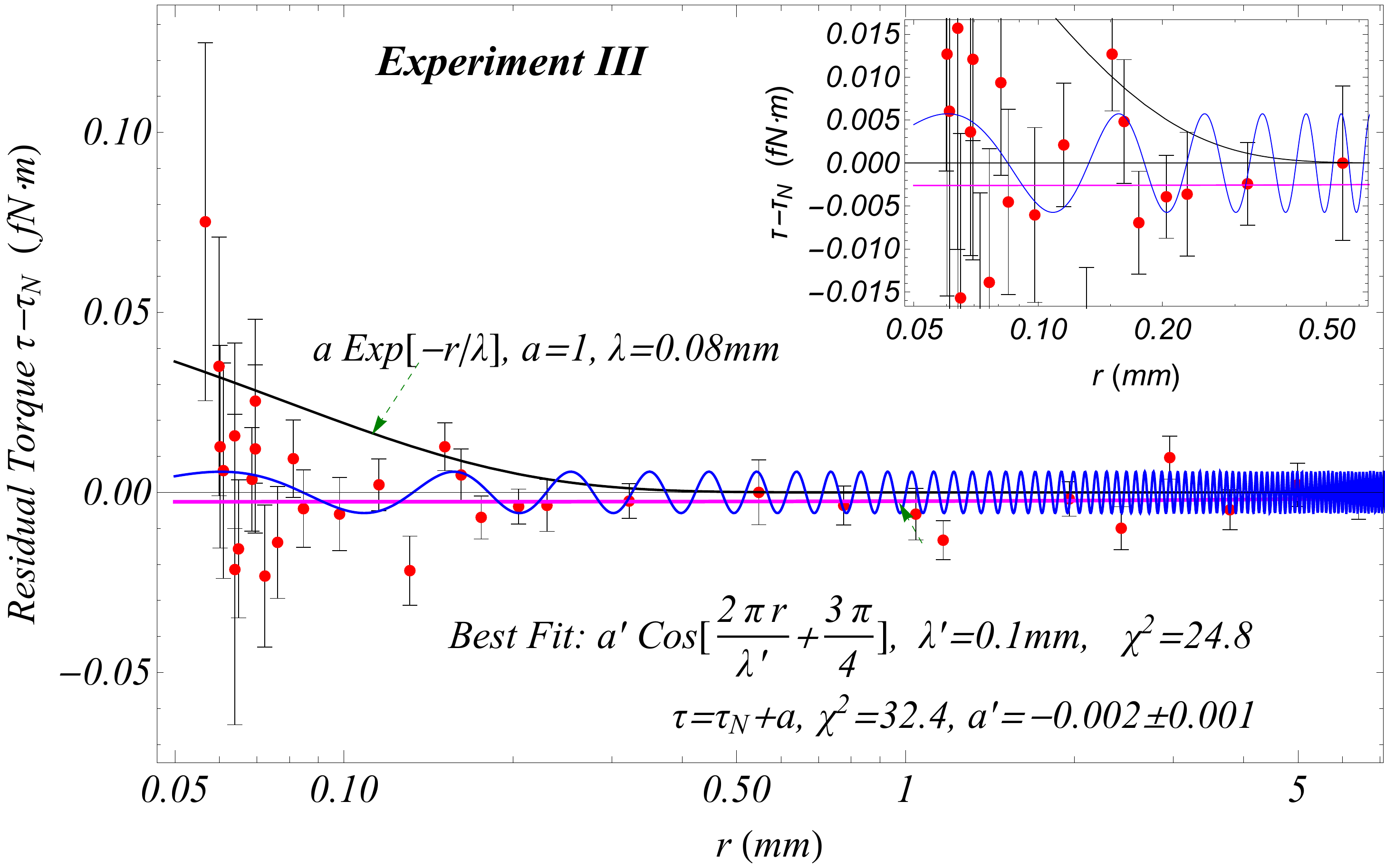} &
\epsfxsize=3.3in
\epsffile{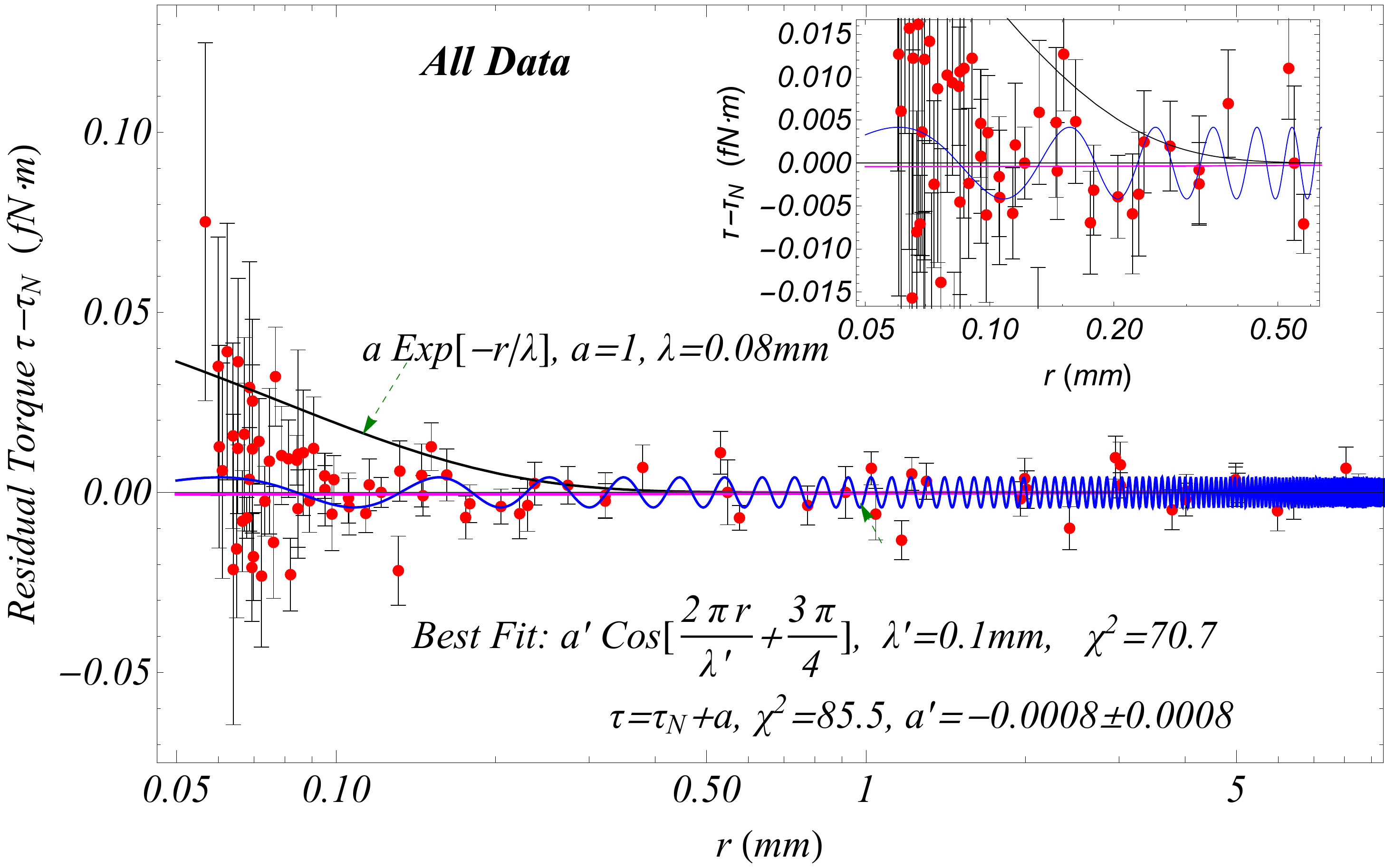}\\
\end{array}$
\end{center}
\vspace{0.0cm}
\caption{\small The residual torque data along with the best fit Yukawa (thick pink line) and oscillating parametrizations (thin blue line). The Newtonian (with offset) best fit value and corresponding $\chi^2$ for each best fit parametrization are also shown. }
\label{dataplot}
\end{figure*}

\begin{center}
\begin{table}[h]
    \begin{tabular}{| c | c |}
    \hline
    {\bf Parametrization} &\;\; $\chi^2$\;\; \\ \hline
   $\delta \tau=\alpha'$ & $ 85.5 $\\ \hline
   $ \delta \tau=\alpha'e^{-m' r}$ & $85.4$\\ \hline
       $ \delta \tau=\alpha'\cos(m' r +\frac{3\pi}{4})$ & $70.7$ \\ \hline   
    \end{tabular}
    \caption[capt1]{The best fit value of $\chi^2$ for each parametrization using the total of $87$ datapoints in the three experiments.}
\label{table:name}
\end{table}
\end{center}

We have fit the residual torques $\delta \tau \equiv \tau-\tau_N$ measured in each experiment for several attractor-detector separations, using three parametrizations:
\ba  
\delta \tau_1(\alpha',m',r) &=& \alpha' \label{constpar} \\
\delta \tau_2(\alpha',m',r) &=& \alpha' e^{-m' r}
\label{exppar} \\
\delta \tau_3(\alpha',m',r) &=& \alpha' \cos(m' r + \frac{3\pi}{4})
\label{oscilpar}
\ea
where $\alpha'$, $m'$ are parameters to be fit. We have fixed $\theta'=\frac{3\pi}{4}$ as it provides better fits than other phase choices. The primes are used to avoid confusion with the corresponding unprimed parameters of the deviations from the Newtonian potential (eg eq. (\ref{ntpot-wash})). 

We have used these parametrizations to minimize $\chi^2(\alpha',m')$ defined as
\be 
\chi^2(\alpha',m')=\sum_{j=1}^N\frac{\left(\delta \tau(j)-\delta \tau_i(\alpha',m',r_j)\right)^2}{\sigma_j^2}
\label{defchi2}
\ee
where $i$ refers to the type of parametrization (\ref{constpar})-(\ref{oscilpar}), $j$ refers to the $j^{th}$ datapoinnt as shown in Table II and $N$ is the number of datapoints in each experiment. The $1\sigma$ and $2\sigma$ contours for two parameters correspond to the curves satisfying $\chi^2(\alpha',m')=\chi_{min}^2 +2.3$ and $\chi^2(\alpha',m')=\chi_{min}^2 +6.17$.

The detailed connection between the parametrization parameters $\alpha', m'$ and $\theta'$ and the corresponding parameters $\alpha, m$ and $\theta$ of the gravitational potential requires detailed knowledge of parameters of the experimental apparatus which are not available to us. These parameters could be used to obtain a quantitative estimate of the source integral and thus of the gravitational signal. An approximate estimate of this signal for the case of macroscopic interacting disks will be obtained in what follows.

In an effort to bypass the calculation of the source integral in the case of the Yukawa parametrization, we have attempted to make an empirical connection between the parametrization parameters and the gravitational signal parameters using the published residual torque curves (with well defined best fit parameters $(\alpha', m')$) that correspond to three pairs of Yukawa potential parameters $(\alpha, m)$. This empirical relation connects $(\alpha', m')$ with the
$(\alpha,m)$ of a Yukawa deviation in the potential and is discussed in detail in the Appendix. It indicates that
\ba 
m'&=&m 
\label{mcon}\\
\ln\left(\frac{\alpha'}{\alpha}\right) &=& 5.65-3.15\; ln(m) \label{alphcon}
\ea
where $m$ is in $mm^{-1}$ and $\alpha'$ in $fN\cdot m$. We stress that these relations are approximately applicable in the case of the Yukawa parametrization and are not necessarily accurate for the derivation of the gravitational signal in the case of the oscillating parametrization. In what follows however we assume that eq. (\ref{mcon}) is applicable also for the oscillating parametrization 
which is justified by the approximate estimate of the source integral discussed below.

In Table I we show the best fit $\chi^2$ values for each parametrization  for the combined dataset from all three experiments. Notice the significant improvement in the quality of fit by $\Delta \chi^2 \simeq -15$ of the oscillating parametrization obtained with $m\simeq 65 mm^{-1}$ which corresponds to a wavelength $\lambda = \frac{2\pi}{m}\simeq 0.1 mm$ (see Fig. \ref{contoursoscil}). This scale is surprisingly close to the dark energy scale $\lambda_{de}\equiv \sqrt[4]{\frac{hc}{\rho_{de}}}=0.085 mm$ as discussed in the Introduction.

The $1\sigma$ and $2\sigma$ contours in the parameter space $(\alpha',m')$ are shown in Fig. \ref{figcontoursexp} (Yukawa parametrization) and in Fig. \ref{contoursoscil} (oscillating parametrization assuming fixed $\theta'=3 \frac{\pi}{4}$) ) for each one of the three experiments and for the combined dataset of 87 datapoints.

Lines of constant $\alpha$ obtained using eq. (\ref{alphcon}) are shown in Fig. \ref{figcontoursexp}). For $\alpha=1$ the line intersects the $2\sigma$ contours at $m=m'\simeq 20 mm^{-1}$ (for the 'all data' plot) thus leading to a $2\sigma$ constraint $m\gsim 20 mm^{-1}$ which is almost identical with the constraint of Ref. \cite{Kapner:2006si-washington3}. This is a good test for the validity of our analysis and of the empirical calibration relations (\ref{alphcon}), (\ref{mcon}).

The value of $\chi^2$ as a function of $m$ for the oscillating parametrization (continous blue line) and for the Newtonian parametrization (straight red line) is shown in Fig. \ref{figchi2vsm} (left panel). For each value of the spatial frequency $m$ we have minimized with respect to the amplitude and the phase of the parametrization. Even though the most prominent $\chi^2$ minimum is  the one at $m\simeq 65 mm^{-1}$ there are some other notable minima. Two of them have a comparable depth with the fundamental deepest minimum at $m=65mm^{-1}$. The spatial frequencies are close to the third harmonic of the fundamental frequency ($m\simeq 195 mm^{-1}$ and $m\simeq 202mm^{-1}$). Even though these two minima would appear to be independent, their corresponding best fit parametrizations behave as higher harmonics of the fundamental minimum. This is demonstrated in Fig. \ref{bfitparams}  where it can be seen that the first few roots of the best fit form of  $m\simeq 202mm^{-1}$ coincide with the corresponding roots of the fundamental best fit at $m=65 mm^{-1}$. 

The effects of horizontal uncertainties on the location and depth of the minima assuming a fixed horizontal error for each datapoint of $\sigma_r=0.002mm$ are shown in Fig. \ref{figchi2vsm} (right panel). In this case $\chi^2$ is evaluated by adding the term $(\frac{\partial \delta \tau}{\partial r} \sigma_r)^2$ in the denominator of the expression defining $\chi^2$ (eq. (\ref{defchi2}). Thus the definition of $\chi^2$ becomes
\be 
\chi^2(\alpha',m')=\sum_{j=1}^N\frac{\left(\delta \tau(j)-\delta \tau(\alpha',m',\theta',r_j)\right)^2}{\sigma_j^2+(\frac{\partial \delta \tau}{\partial r} \sigma_r)^2}
\label{defchi2s}
\ee
As shown in Fig. \ref{figchi2vsm} (right panel), the location of the minima is not affected by this introduction of horizontal errors even though the depth of the higher frequency minima appears to increase. The depth and location of the fundamental minimum at $m=65 mm^{-1}$ are practically unaffected by taking into account such horizontal errors since the relevant scale of these errors is much smaller.

The residual torque data along with the best fit Yukawa and oscillating parametrizations are shown in Fig. \ref{dataplot} for each experiment separately as well as for the combined dataset. The values of $\chi^2$ for the best fit and for the corresponding constant fit (Newtonian plus constant offset) are also shown on each plot.

\begin{figure}[!t]
\centering
\vspace{0cm}\rotatebox{0}{\vspace{0cm}\hspace{0cm}\resizebox{0.49\textwidth}{!}{\includegraphics{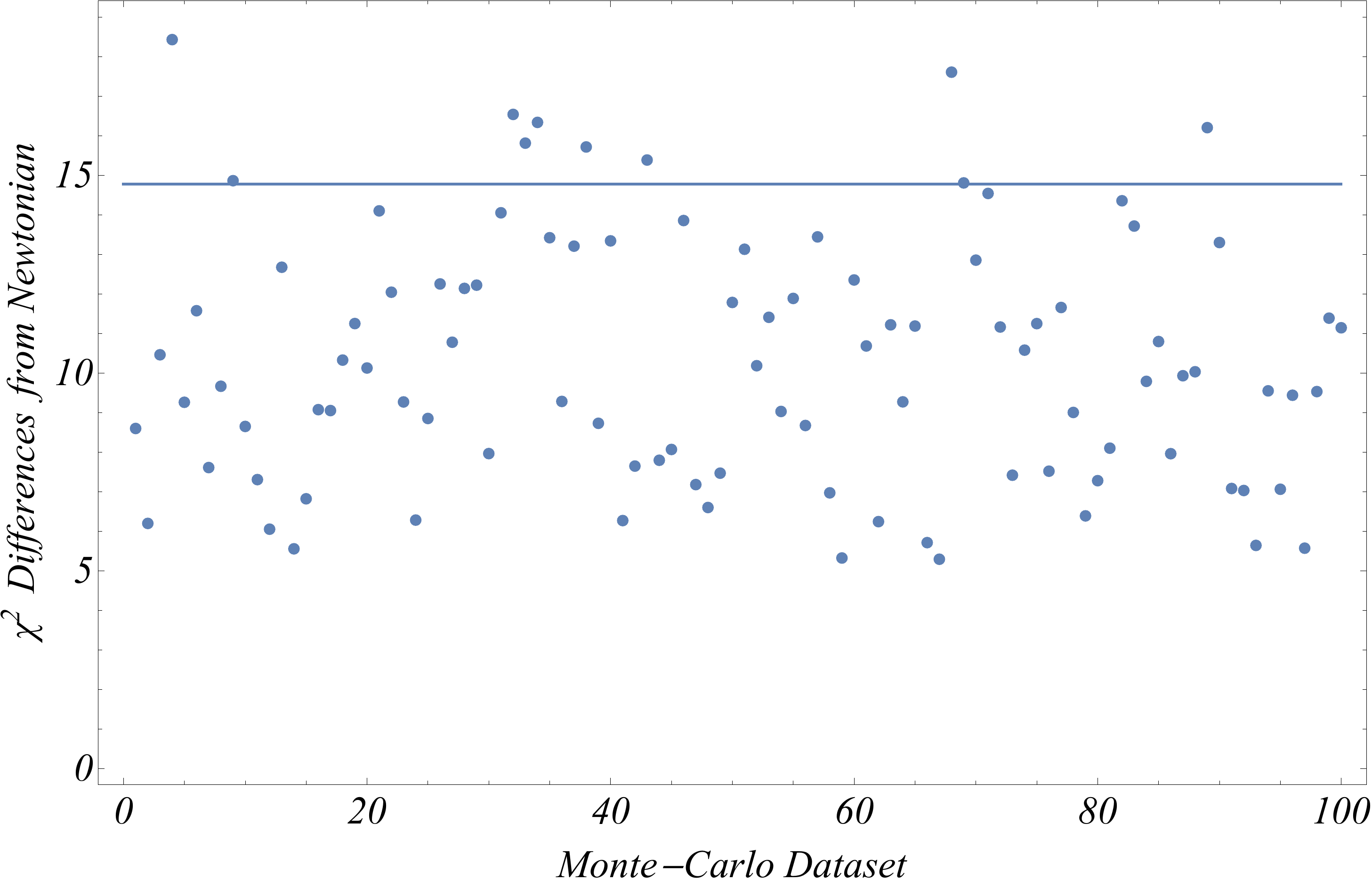}}}
\caption{The $\chi^2$ differences between the deepest $\chi^2$ in the range $m\in[0,100]$ and the corresponding Newtonian value of $\chi^2$ in 100 Monte Carlo datasets created under the assumption of zero residuals (Newtonian model). }
\label{montecarlofits}
\end{figure}

In view of the presence of the additional minima of $\chi^2$ shown in Fig. \ref{figchi2vsm} it becomes clear that the $3\sigma$ level of significance of the $m=65mm^{-1}$ minimum that emerges from the contour plot of Fig. \ref{contoursoscil} is a overestimate. In order to obtain a better estimate of the level of significance of this minimum we have constructed 100 Monte Carlo realizations of the Washington data assuming an underlying Newtonian model. We used the same errorbars and $r$ coordinates of the original data and assumed a Gaussian distribution for each datapoint around a zero mean with standard deviation equal to the data errorbars. We then fit these datasets with oscillating parametrizations in the range $0-100mm^{-1}$ which includes the identified fundamental frequency $65 mm^{-1}$ by varying both the amplitude and the phase of the parametrization for each value of the spatial frequency $m$. We found that about 10\% of the deepest minima of the  Monte Carlo datasets are deeper than the observed fundamental minimum at $m=65 mm^{-1}$ (see Fig. \ref{montecarlofits}). Thus, even though the oscillating parametrization considered is a viable fit to the data, the level of significance of the corresponding $\chi^2$ minimum is not more than $2\sigma$.

\subsubsection{Source Integral: An estimate of the expected gravitational signal}
\label{sec:Section 322}

Our assumption of a harmonic parametrization for the fit of the torque residual data is a simplified approximation based on the theoretically predicted oscillating force forms of eqs. (\ref{forcenonlocal}) and (\ref{forcefR}) between point particles. In the case of realistic forces between macroscopic bodies, the predicted interaction force is expected to be modified. In the case of the Washington experiments, the relevant macroscopic bodies are disks of approximate radius of $2.5mm$ and thickness about $1mm$ corresponding to the missing mass holes of the apparatus.

We have obtained an independent numerical approximate estimate of the residual force that would be present between two disks in the presence of harmonic spatial oscillations of the Newtonian potential. In this calculation we have assumed a modified Newtonian force field motivated by non-local gravity (eg. (\ref{forcenonlocal})), discretized each disk to a grid with large number of segments of scale $\Delta x$. Assuming two disks of the same radius $R$ with symmetry axes parallel to the $z$ axis and centers at the origin and at $(x_0,y_0,z_0)$ respectively, we discretize the radius as $R={\bar R} \Delta x$ and the height as $2h=2{\bar h} \Delta x$ where ${\bar R}$ and ${\bar h}$ are integers denoting the dimensions of the disks in units of $\Delta x$. The Newtonian force between two cubic segments with central coordinates $(x_i,y_i,z_i)$ (included in the disk centered at the origin) and $(x_j,y_j,z_j)$ (included in the disk centered at ($(x_0,y_0,z_0)$) is of the form
\begin{widetext}
\ba
{\vec F}_{ij}^N&=&- G \rho_1 \rho_2 (\Delta x)^6 \frac{{\vec r}_j-{\vec r}_i}{((x_j-x_i)^2+(y_j-y_i)^2+(z_j-z_i)^2)^{3/2}}\nn \\
&=&- G \rho_1 \rho_2 (\Delta x)^4 \frac{{\vec {\bar r}}_j-{\vec {\bar r}}_i}{((\bar x_j-\bar x_i)^2+(\bar y_j-\bar y_i)^2+(\bar z_j-\bar z_i)^2)^{3/2}} \nn \\
&\equiv&- G \rho_1 \rho_2 (\Delta x)^4 \; \frac{{\vec {\bar r}}_{ij}}{{\bar r}_{ij}^3}
\label{fij}
\ea
\end{widetext}

\begin{figure*}[ht]
\centering
\begin{center}
$\begin{array}{@{\hspace{-0.10in}}c@{\hspace{0.0in}}c}
\multicolumn{1}{l}{\mbox{}} &
\multicolumn{1}{l}{\mbox{}} \\ [-0.2in]
\epsfxsize=3.3in
\epsffile{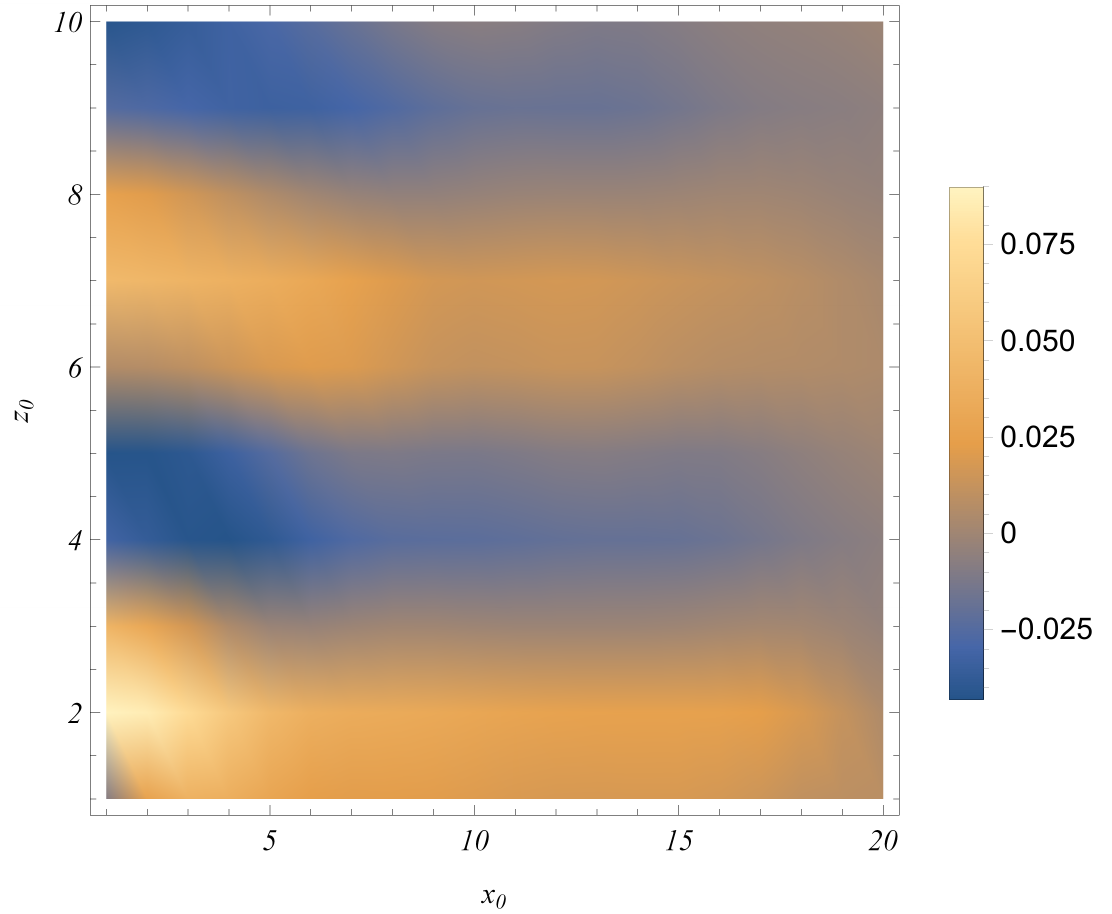} &
\epsfxsize=3.3in
\epsffile{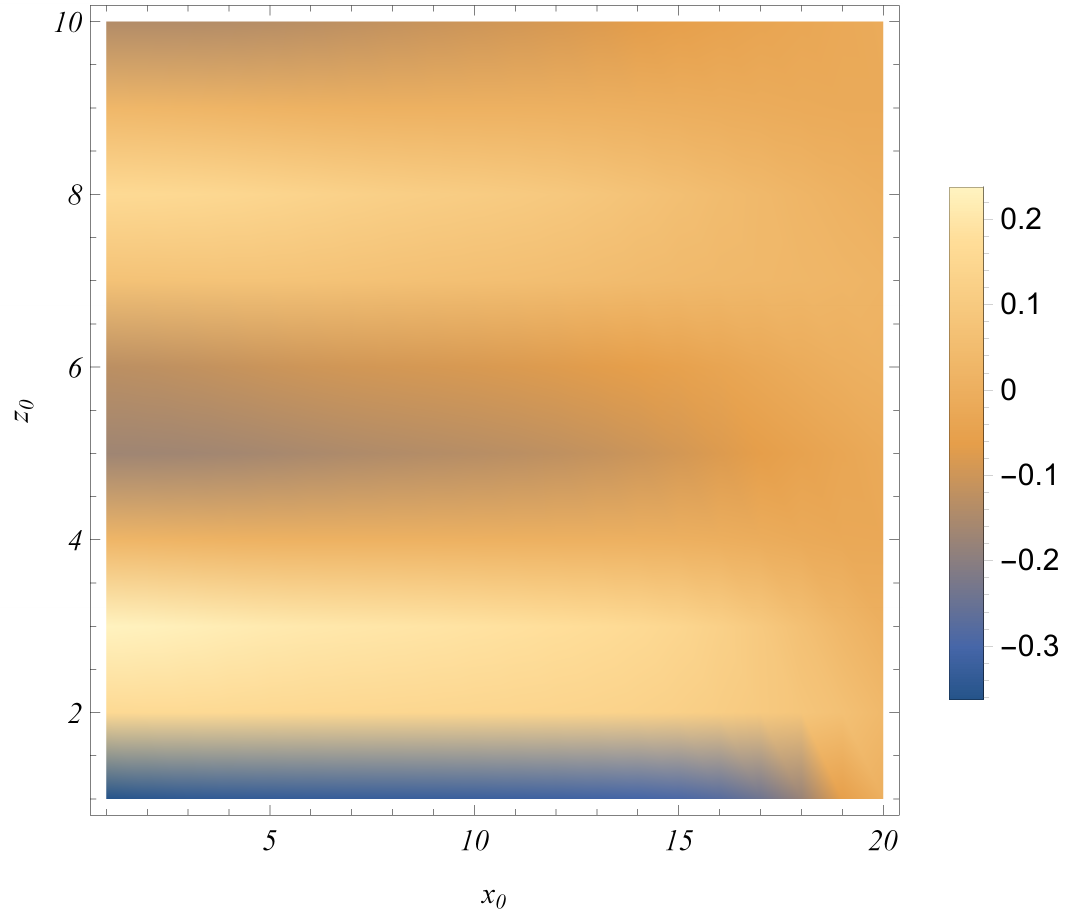} \\
\end{array}$
\end{center}
\vspace{0.0cm}
\caption{\small The residual x (left) and z (right) dimensionless force components for disk radius $R=10 \Delta x$ and oscillation wavelength $\lambda=5 \Delta x$. The axes units are $\Delta x$.}
\label{figsresdp}
\end{figure*}

\begin{figure*}[ht]
\centering
\begin{center}
$\begin{array}{@{\hspace{-0.10in}}c@{\hspace{0.0in}}c}
\multicolumn{1}{l}{\mbox{}} &
\multicolumn{1}{l}{\mbox{}} \\ [-0.2in]
\epsfxsize=3.3in
\epsffile{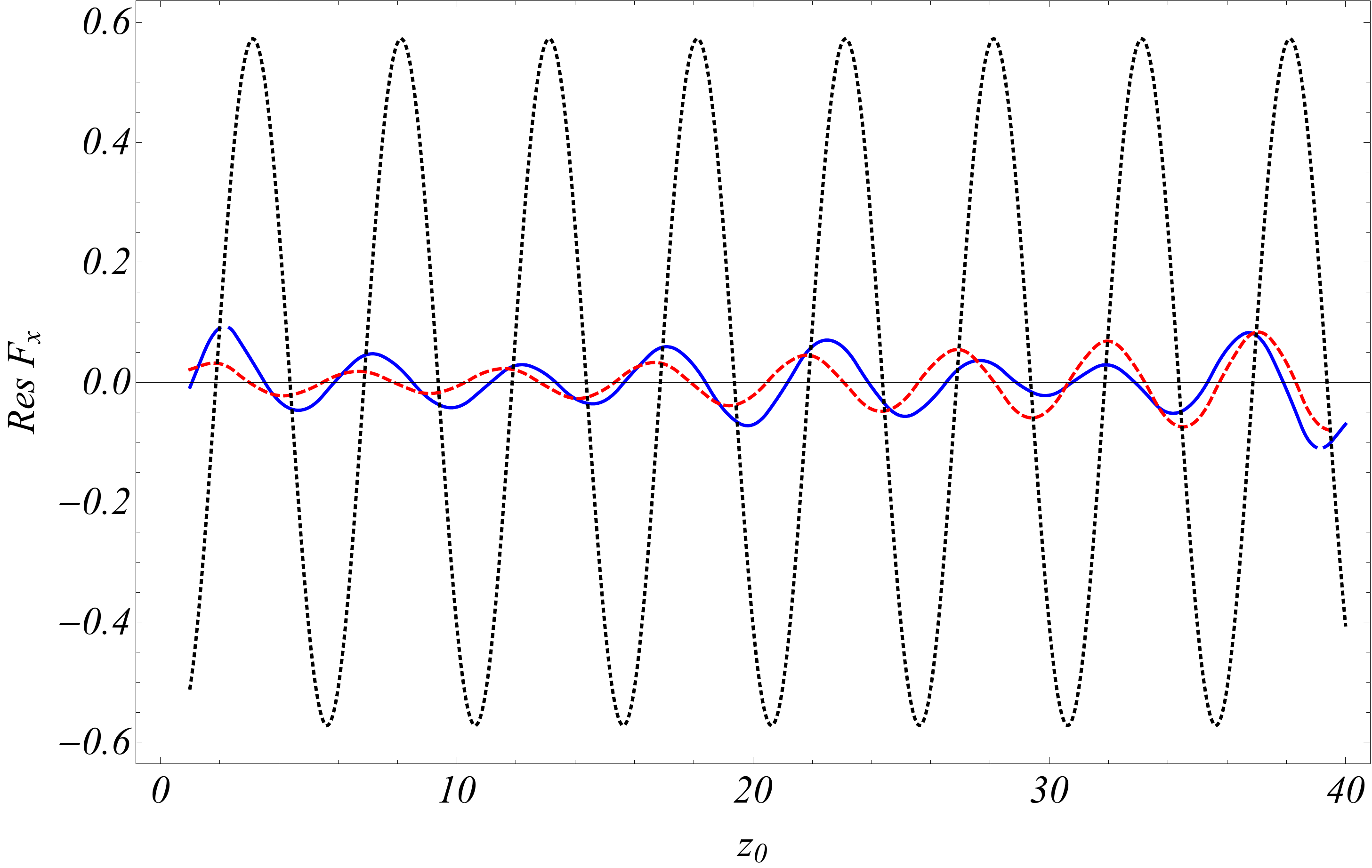} &
\epsfxsize=3.3in
\epsffile{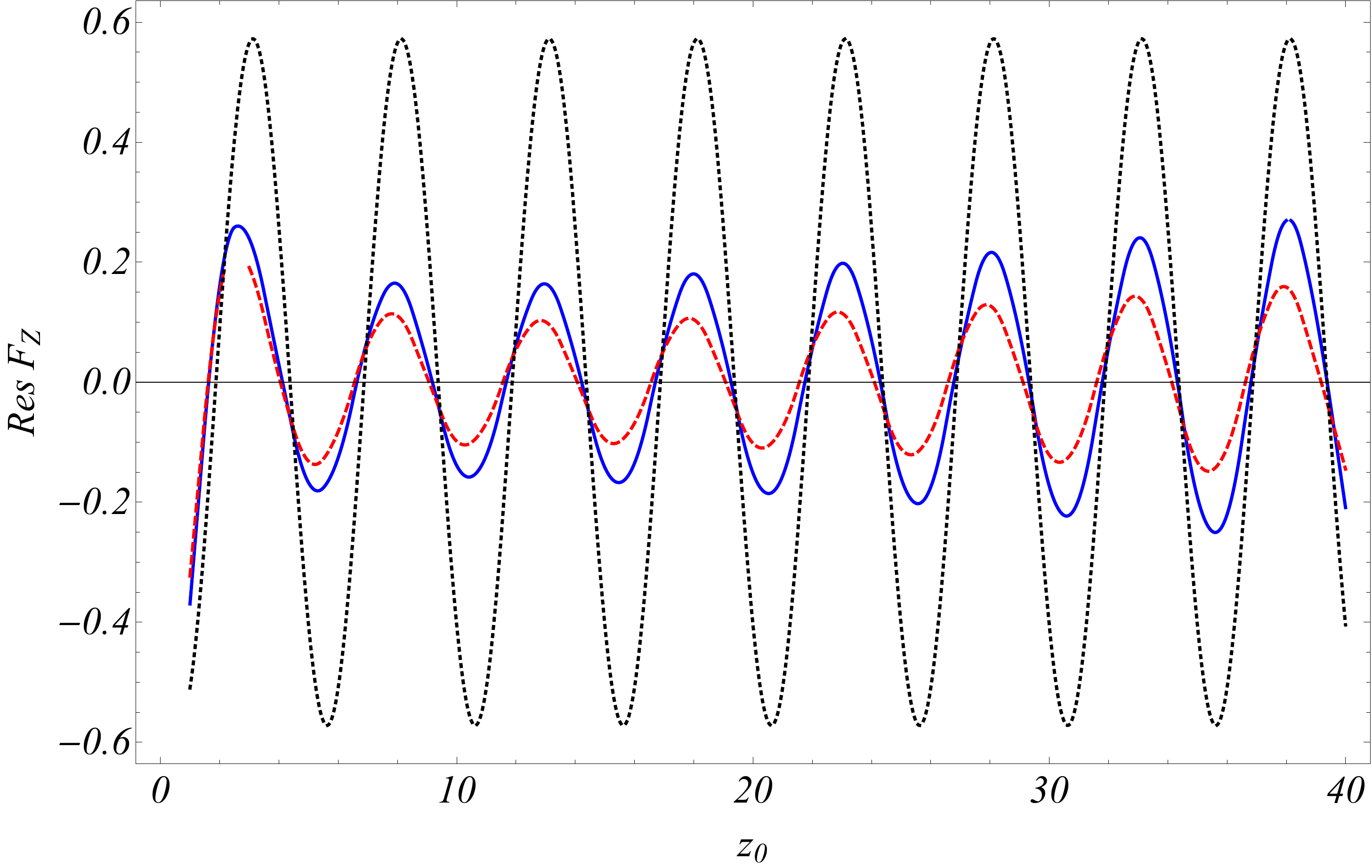} \\
\end{array}$
\end{center}
\vspace{0.0cm}
\caption{\small The residual x (left) and z (right) dimensionless force components for disk radius $R=10 \Delta x$, $\lambda=5 \Delta x$. The axes units are $\Delta x$. The blue line for the x-component (left panel) corresponds to $x_0=1$ while the red line to $x_0=10$. The blue line for the z-component (right panel) corresponds to $x_0=0$ while the red line to $x_0=10$. The black line corresponds to the naively expected signal $a \cos(0.5 z_0 + \theta)$ (we used $\theta =3\pi/4$ for the z-component which seems to fit both the theoretically predicted and the observed signal). Notice the predicted suppression of the signal at low distances $z_0$ and its slow  increase of the predicted residual with distance. It may be shown that the amplitude grows  up to the maximum value of $0.57$ as expected since at large distances the disks behave as pointlike particles.}
\label{figsres}
\end{figure*}

where $\rho_1$, $\rho_2$ are the disk densities and the `barred' quantities are integers such that eg $x_i={\bar x}_i \Delta x_i$. The total force between the disks is  simply approximated as
\be
{\vec F}_{tot}^N=\sum_{z_{imin}}^{z_{imax}}\sum_{z_{jmin}}^{z_{jmax}}\sum_{x_{imin}}^{x_{imax}}\sum_{y_{imin}}^{y_{imax}}\sum_{x_{jmin}}^{x_{jmax}}\sum_{y_{jmin}}^{y_{jmax}} {\vec F}_{ij}^N
\ee
where the summation limits corresponding to  the pair of disks described above are obtained as (in what follows we omit the `bars')
\ba
z_{imin}&=&-h \\
z_{imax}&=&h \\
x_{imin}&=&-R \\
x_{imax}&=&R \\
y_{imin}&=&-\sqrt{R^2-x_i^2} \\
y_{imax}&=&\sqrt{R^2+x_i^2} \\
z_{jmin}&=&z_0-h \\
z_{jmax}&=&z_0+h \\
x_{jmin}&=&x_0-R \\
x_{jmax}&=&x_0+R \\
y_{jmin}&=&y_0-\sqrt{R^2-(x_0-x_i)^2} \\
y_{imax}&=&y_0+\sqrt{R^2+(x_0-x_i)^2} 
\ea
The generalized gravitational force predicted by nonlocal gravity theories is nonzero only for $m\; r>1$ and it is of the form
\begin{widetext}
\ba
\vec{F}^O&=&-\hat{r}\frac{G m_i m_j}{r^2}\left(1+\frac{2\alpha_2 \cos(m r +\theta)}{m r}+\alpha_2 \sin(m r +\theta)\right) \\
&=& - G \rho_1 \rho_2 (\Delta x)^4 \; \frac{{\vec r}_{ij}}
{r_{ij}^3} 
\left(1+\frac{2\alpha_2 \cos(m r_{ij} +\theta)}{m r_{ij}}
+ \alpha_2 \sin(m r_{ij} +\theta)\right)
\label{forcenonlocal1}
\ea
\end{widetext}
where $m=\frac{2\pi}{\lambda}$ is the fundamental scale of the theory, and the theoretically predicted parameter values were evaluated in section II as $\alpha_2=0.572$ and $\theta=0.885\pi$.
The predicted residual $x$-component of the force for this class of models may be obtained in dimensionless form as
\be
F_{xRes}\equiv \frac{F_x^O-F_x^N}{F_x^N}
\label{resx}
\ee
and similarly for the $y$ and $z$ components. Using these definitions we  evaluate $F_{xRes}$ and $F_{zRes}$ as functions of the center of the second disk coordinates $z_0$ and $x_0$ for $y_0=0$. We have set $R=10$, $\lambda=5$ and $h=0$. The assumed unit is $\Delta x$. For example for $\lambda = 100\mu m$ we would have $\Delta x=20\mu m$ and $R=200\mu m$. 

For discs with radius of $R=2mm$ and thickness $2h=1mm$,  we would have to setup a grid with  $R=100$ and $h=25$ for the same value of $\Delta x$ which is needed for proper probing of the oscillations (at least $5$ grid points per spatial oscillation wavelength). This calculation would take $5000^2$ times longer to run than the calculation for the parameters used above which makes such choice impractical. The above implemented choice of dimensions however provides also fairly useful insight about the predicted signal. 

The choice of disk parameters used, corresponds to division of each disk to a grid with more than 300 segments while thickness is ignored. The assumed spatial wavelength of the oscillating potential was half the radius of the disk. These assumptions allowed this simple code to give results in relatively short time for a wide range of relative positions between the disks. These results are shown in Figs \ref{figsresdp} and \ref{figsres}. In Fig. \ref{figsresdp} we show a density plot of the predicted residual force x and z components while in Fig. \ref{figsres} we show the same components when the second disk is placed along particular spatial directions parallel to the axes of the disks (z direction). Despite of the qualitative nature of this approach it led to four interesting conclusions
\begin{enumerate}
\item
 The predicted signal for the force between macroscopic bodies remains oscillatory but for distances comparable to the disk dimensions, it is not a harmonic function with constant amplitude. 
\item
The amplitude of the oscillating force on scales smaller than the macroscopic bodies is suppressed compared to the amplitude of the fundamental potential oscillations.
\item
The spatial frequency of the macroscopic force is about the same as the frequency of the fundamental oscillations.
\item
The magnitude of the residual of the oscillating force defined in eq. (\ref{resx}),  tends to increase slowly with distance. We have verified that the magnitude of the oscillating force tends to its pointlike value ($a_2=0.572$) at large distances between the disks as expected. As shown in Figs \ref{figsresdp} and \ref{figsres}, the signal also exists off axis even though it is reduced compared to the signal when the symmetry axes of the two disks coincide.
\end{enumerate}
Based on the above conclusions, an experiment designed to detect spatial oscillations of the gravitational potential should focus on relatively large source-detector distances but vary these distances by small steps comparable to the targeted oscillation wavelength. In this manner the relative residual magnitude of the signal is maximized while the spatial frequency of the signal is properly probed.

\section{Conclusions-Discussion}
\label{sec:Section 4}

We have shown that the presence of sub-millimeter oscillations of the gravitational potential is a viable possibility at both the theoretical and the observational/experimental level. 

At the theoretical level we showed that gravitational potential oscillations appear generically in stable extended gravitational theories like non-local ghost free gravity. Even in theories where such oscillations are generic but unstable we showed that there is potential for stabilization mechanisms induced by nonperturbative effects. 

At the macroscopic observational level we showed that even though these spatial oscillations do not have a strict Newtonian limit, they are consistent with observations and gravitational experiments for small enough value of the wavelength. In fact we presented evidence that such oscillating parametrizations may provide a better fit to torsion balance data than the Newtonian potential parametrization even though the level of significance of this improvement is not more than $2\sigma$.  Our data analysis involved several assumptions and simplifications especially in the estimate of the source integral. The goal of our data analysis has not been the quantitative estimate of a statistically significant oscillating signature of modified gravity on sub-milimeter scales. Instead, it has been the demonstration of the existence of a peculiar oscillating  signal in the data which may be statistically significant. This signal could be due to three possible effects:
\begin{itemize}
\item
A statistical fluctuation of the data which is more prominent in Experiment III of \cite{Kapner:2006si-washington3} as shown in Fig. \ref{contoursoscil}.
\item
A periodic distance-dependent systematic feature in the data. Such an unnoticed class of systematics would not be of wide interest but it would be notable and useful for the short-distance force measurement community\footnote{I thank the referee for pointing this out}.
\item
An early signal for a short distance modification of GR. The verification of this possibility would require extensive detailed search of short-distance force measurement groups which will hopefully be motivated by our analysis. Other indications of such oscillating short-distance forces may be seen by eye on the recent data plots (Fig. 3) of Ref. \cite{Rider:2016xaq-recent-constr-oscillations-evident}.
\end{itemize} 

In view of the possible future verification of the last possibility, the following challenging question needs to be addressed 'What is the precise form and amplitude of the theoretically predicted oscillating signal in the context of the Washington experiment?'. In order to address this question, a calculation of the torque between the pendulum disk and upper attractor for several separations is needed. In the present analysis we have made some progress for addressing this question at a qualitative level (section \ref{sec:Section 322}) but the detailed quantitative answer is a challenging issue that remains open and will be addressed in a separate forthcoming analysis. The challenging nature of this question emerges not only because of the small grid spacing required compared to the oscillation wavelength  (much larger grid required) but also due to the nonlocal nature of the signal that requires the lower attractor of the Washington experiment to be included in any full calculation.

Our results may have a few interesting implications:
\begin{itemize}
\item
They may be viewed as early evidence for emerging signatures of non-local gravity in experimental data. 
The idea of non-local gravity, provides one of
the very few self consistent approaches for curing the singularity problems
of General Relativity while being naturally motivated by the demand of
consistency with quantum mechanics. It is remarkable that such a well
motivated extension of General Relativity generically predicts the presence
of sub-millimeter spatial oscillations of Newton's constant. The prediction
for these oscillations was documented clearly in Ref. \cite{Edholm:2016hbt-nonlocal-potential-stable-spatial-oscillations} and was also seen in Refs \cite{Kehagias:2014sda-nonlocal-oscillations,Maggiore:2014sia-nonlocal-gravity-oscil}.
\item
They indicate that oscillating parametrizations for deviations from the the Newtonian gravitational potential should be considered along with Yukawa and power law parametrizations because they are well motivated theoretically and consistent with macroscopic observations.
\item
They indicate that the stability of $f(R)$ theories that are unstable at the perturbative level could be re-examined by taking into account non-trivial backgrounds and the backreaction from non-linear terms that act at the non-perturbative level.
\end{itemize}
Interesting extensions of the present work include the following:
\begin{itemize}
\item
A more detailed analysis of the existence and stability of theoretical models that predict the existence of sub-millimeter oscillations of the gravitational potential and of Newton's constant. In addition to the models discussed in the present analysis such models may include the presence of compact time-like extra dimensions \cite{Quiros:2007ym-timelike-extra-dim}, brane-world models etc.
\item
A detailed analysis of the macroscopic effects of sub-millimeter oscillations of Newton's constant including possible effects on solar system scales and/or Lunar Ranging experiments.
\item
The cosmological effects of such oscillations could also be of significant interest especially in view of the experimental indications that they may exist with a wavelength close to the dark energy scale. For example oscillations of Newton's constant in time are generically present in stable scalar tensor theories and theories with compact extra dimensions and have been shown \cite{Steinhardt:1994vs-temporal-oscillations,Perivolaropoulos:2003we-temporal-oscillations} to have interesting cosmological features including their possible role as dark energy candidates.
\item
The use of alternative parametrizations to fit the torque-residual data. It may be possible to find alternative parametrizations that provide better fits which may provide hints for the construction of new theoretical models. It is also important however  to identify possible sources of systematics in the data that may induce spurious non-physical features. For example the minor systematic effect in Experiment I coming from the slightly bowed detector ring could be the origin of the difference of best fit spatial wavelength by a factor of 2 found in the data analysis of Experiment I.
\item
If the observed signal in the Washington group data is physically interesting it should leave a signature on other experiments which have the required sensitivity to see the signal (eg \cite{Rider:2016xaq-recent-constr-oscillations-evident,Yang:2012zzb-recent-experiment,Tan:2016vwu-test-submm-newton-law-less-constraining,Cook2013-thesis,Hagedorn2015-thesis,Kapner:2005qy-thesis}). It is therefore important to extend this analysis to other datasets in an effort to identify similar signatures in other datasets. An investigation along these lines is currently in progress.
\end{itemize}
In conclusion, we have presented a novel and potentially
important result that could motivate further work in the active field of theoretical and experimental searches for modification of GR.

{\bf Numerical Analysis:} The Mathematica file that led to the production of the figures may be downloaded from \href{http://leandros.physics.uoi.gr/newt-oscil}{here}.

\section*{Acknowledgements}
I thank Savvas Nesseris for his help with the data analysis. I also thank the referees for useful comments and especially referee B (experimentalist) for taking the time to reproduce in detail and extend the residual data analysis presented here.

\section*{Appendix}

In this Appendix we describe the derivation of the empirical relations (\ref{mcon}) and (\ref{alphcon}) and we show the residual torque  data used for the fit of the parametrizations considered.

\begin{figure}[!t]
\centering
\vspace{0cm}\rotatebox{0}{\vspace{0cm}\hspace{0cm}\resizebox{0.49\textwidth}{!}{\includegraphics{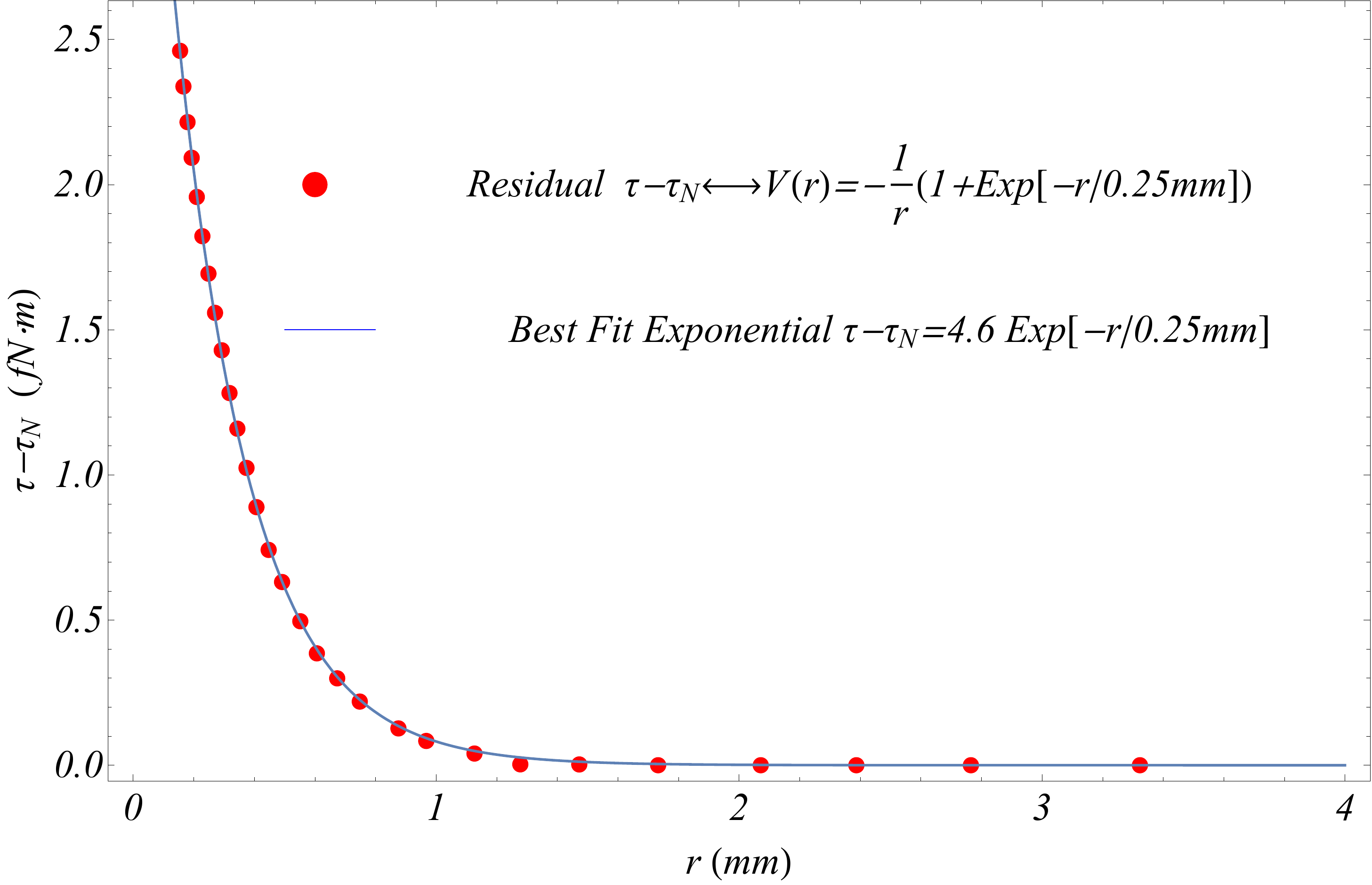}}}
\caption{The curve describing a Yukawa deviation from the Newtonian potential in the torque residuals (thick dots) is fit well as an exponential with the same value of $m$ (continous line).}
\label{expfitcalib}
\end{figure}

In order to calibrate the measured torque residuals and connect them with the parameters of the deviations from the Newtonian potential ($\alpha$ and $m$ of eq. (\ref{ntpot-wash})) we use the published residual torque curves that correspond to Yukawa type Newtonian deviations for specific parameter values. We find that these residual curves are very well fit by exponentials with the same exponents as the exponents of the Yukawa Newtonian potential deviations. This is consistent with eq. (\ref{ntforceexp1}) which implies that the dominant residual torque for $m r>1$ is 
\be 
\tau-\tau_N \sim e^{-m r}
\label{restoerquepred}
\ee
This exponential behavior of the residual torque is verified in Fig. \ref{expfitcalib} where we show the published in Ref. \citep{Hoyle:2000cv-washington1} form of the residual torque (thick dots) corresponding to a Yukawa deviation with $\alpha=1$ and $\lambda\equiv m^{-1}= 0.25mm$ superposed with the best fit exponential $\alpha' e^{-r/\lambda'}$. As expected we find excellent fit for $\lambda'=\lambda=0.25mm$. Also since $\alpha=1$ and the best fit is $\alpha'=4.6$ we have $\frac{\alpha'}{\alpha}=4.6$. 

\begin{figure*}[ht]
\centering
\begin{center}
$\begin{array}{@{\hspace{-0.10in}}c@{\hspace{0.0in}}c}
\multicolumn{1}{l}{\mbox{}} &
\multicolumn{1}{l}{\mbox{}} \\ [-0.2in]
\epsfxsize=3.3in
\epsffile{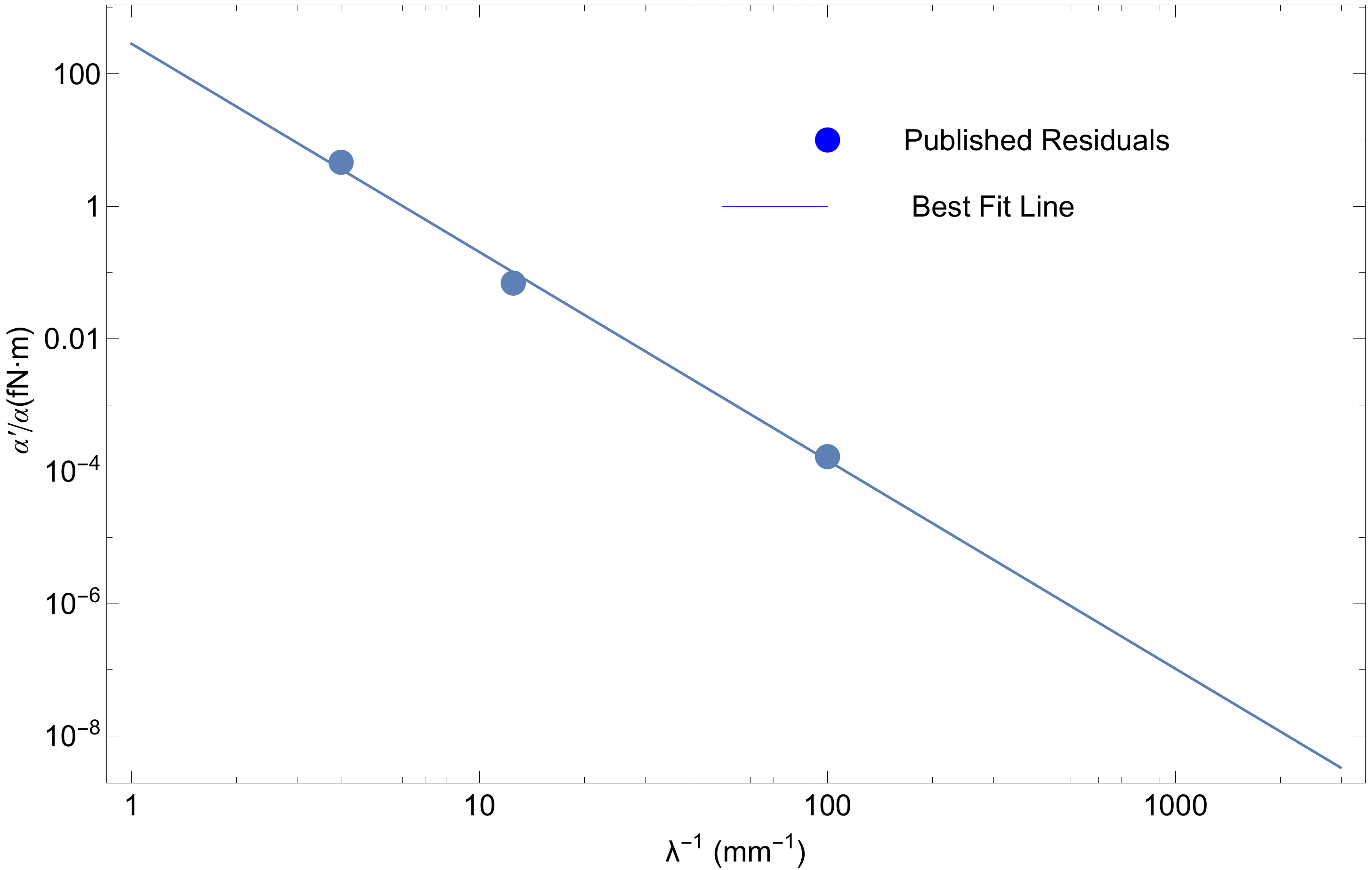} &
\epsfxsize=3.3in
\epsffile{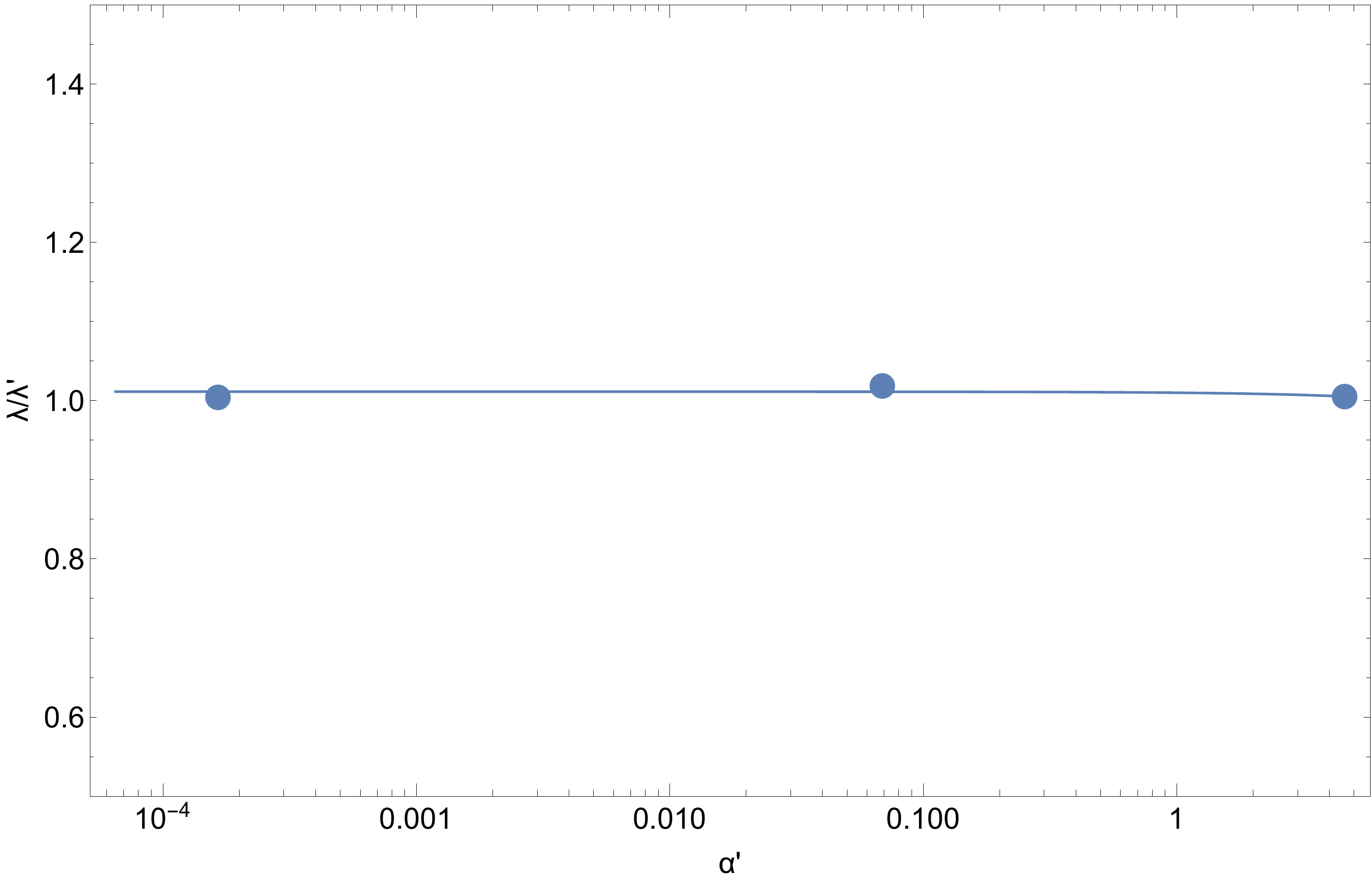} \\
\end{array}$
\end{center}
\vspace{0.0cm}
\caption{\small Using three plots like the one shown in Fig. \ref{expfitcalib} we may relate the parameters $(\alpha,m)$ of the potential deviations with the corresponding parameters $(\alpha',m')$ in the space of torque residuals fit with the same parametrization. These plots demonstrate the validity of eqs (\ref{mcon}) and (\ref{alphcon}) (continous lines through points). }
\label{figparcon}
\end{figure*}

Using also the other two similar residual torque curves of Ref. \citep{Kapner:2006si-washington3} for $(\alpha,m^{-1})=(1,0.08mm)$ and $(\alpha,m^{-1})=(10^5,0.01mm)$,  we evaluate the corresponding ratios $\frac{\lambda}{\lambda'}$ and $\frac{\alpha'}{\alpha}$ and thus construct Fig. \ref{figparcon}. Using a proper fit to these points we derive the empirical relations (\ref{alphcon}), (\ref{mcon}) which relate $(\alpha,m)$, with $(\alpha',m')$.

Finally, in Table II we show the datapoints used to find the best fit for the three parametrizations (\ref{constpar}), (\ref{exppar}) and (\ref{oscilpar}). These datapoints, obtained from Figs. 3, 4 and 5 of Ref. \citep{Kapner:2006si-washington3} using plot-digitizer software\footnote{http://arohatgi.info/WebPlotDigitizer/}, are shown in Table II. The uncertainties in the measurement of $r$ in the
data of Table II is in the range of $0.002-0.005mm$ which is much smaller
than the scale of the oscillation signal. As shown in Fig. \ref{figchi2vsm} the uncertainties of $r$ in this range leave the $\chi^2$ minima in the range around $m\in [0,100]mm^{-1}$ practically unaffected.


\begin{longtable}{ | c | c | c | c |}
\caption{The residual torque 87 datapoints used for the $\chi^2$ analysis.}\label{tab:b}\\
\hline
    $r \; mm$ &  $\tau-\tau_N$ ($fN\cdot m$)& $1\sigma \; (\tau-\tau_N)$   & Experiment \\ \hline
 0.062 & 0.039 & 0.036 & I \\
 0.065 & 0.036 & 0.023  & I\\
 0.067 & -0.008 & 0.014  & I\\
 0.068 & -0.007 & 0.006  & I\\
 0.07 & -0.018 & 0.012  & I\\
 0.073 & -0.002 & 0.01  & I\\
 0.077 & 0.032 & 0.014  & I\\
 0.084 & 0.009 & 0.007  & I\\
 0.095 & 0.005 & 0.006  & I\\
 0.106 & -0.004 & 0.008  & I\\
 0.114 & -0.006 & 0.005  & I\\
 0.146 & -0.001 & 0.006  & I\\
 0.237 & 0.002 & 0.006  & I\\
 0.379 & 0.007 & 0.006  & I\\
 0.577 & -0.007 & 0.003  & I\\
 0.915 & 0. & 0.007  & I\\
 1.301 & 0.003 & 0.005  & I\\
 1.995 & 0.004 & 0.006  & I\\
 3.021 & 0.008 & 0.006  & I\\
 4.027 & 0. & 0.005  & I\\
 5.04 & 0.001 & 0.004  & I\\
 8.512 & 0.001 & 0.004  & I\\
 0.065 & 0.012 & 0.018  & II\\
 0.067 & 0.016 & 0.027  & II\\
 0.069 & 0.029 & 0.035  & II\\
 0.069 & -0.021 & 0.015  & II\\
 0.072 & 0.014 & 0.012  & II\\
 0.075 & 0.009 & 0.02  & II\\
 0.079 & 0.01 & 0.014 & II\\
 0.082 & -0.023 & 0.01 & II\\
 0.085 & 0.011 & 0.029 & II\\
 0.087 & 0.011 & 0.017 & II\\
 0.089 & -0.002 & 0.009 & II\\
 0.091 & 0.012 & 0.014 & II\\
 0.095 & 0.001 & 0.01 & II\\
 0.095 & 0.001 & 0.007 & II\\
 0.099 & 0.004 & 0.007 & II\\
 0.106 & -0.002 & 0.007 & II\\
 0.122 & 0. & 0.004 & II\\
 0.132 & 0.006 & 0.008 & II\\
 0.145 & 0.005 & 0.009 & II\\
 0.179 & -0.003 & 0.005 & II\\
 0.222 & -0.006 & 0.007 & II\\
 0.274 & 0.002 & 0.005 & II\\
 0.322 & -0.001 & 0.006 & II\\
 0.531 & 0.011 & 0.006 & II\\
 1.024 & 0.007 & 0.005 & II\\
 1.221 & 0.005 & 0.005 & II\\
 2.014 & 0.001 & 0.005 & II\\
 3.021 & 0.002 & 0.003 & II\\
 4.014 & -0.002 & 0.005 & II\\
 4.983 & 0.004 & 0.004 & II\\
 5.981 & -0.005 & 0.006  & II\\
 8.054 & 0.007 & 0.006 & II\\
 0.057 & 0.075 & 0.05 & III\\
 0.06 & 0.035 & 0.036 & III\\
 0.06 & 0.013 & 0.028 & III\\
 0.061 & 0.006 & 0.03 & III\\
 0.064 & 0.016 & 0.026 & III\\
 0.064 & -0.021 & 0.043 & III\\
 0.065 & -0.016 & 0.019 & III\\
 0.069 & 0.004 & 0.014 & III\\
 0.07 & 0.012 & 0.023 & III\\
 0.07 & 0.025 & 0.023 & III\\
 0.072 & -0.023 & 0.02 & III\\
 0.076 & -0.014 & 0.016 & III\\
 0.081 & 0.009 & 0.011 & III\\
 0.085 & -0.005 & 0.011 & III\\
 0.098 & -0.006 & 0.01 & III\\
 0.116 & 0.002 & 0.007 & III\\
 0.131 & -0.022 & 0.01 & III\\
 0.151 & 0.013 & 0.007 & III\\
 0.162 & 0.005 & 0.007 & III\\
 0.176 & -0.007 & 0.006 & III\\
 0.205 & -0.004 & 0.005 & III\\
 0.23 & -0.004 & 0.007 & III\\
 0.322 & -0.002 & 0.005 & III\\
 0.548 & 0. & 0.009 & III\\
 0.777 & -0.004 & 0.005 & III\\
 1.044 & -0.006 & 0.007 & III\\
 1.168 & -0.013 & 0.005 & III\\
 1.958 & -0.002 & 0.005 & III\\
 2.423 & -0.01 & 0.006 & III\\
 2.957 & 0.01 & 0.006 & III\\
 3.784 & -0.005 & 0.006 & III\\
 4.993 & 0.002 & 0.006 & III\\
 6.416 & -0.002 & 0.005 & III\\    
\hline
\end{longtable}

\raggedleft
\bibliography{bibliography}

\end{document}